\documentclass[fleqn,usenatbib]{mnras}

\usepackage{newtxtext,newtxmath}
\usepackage[T1]{fontenc}
\usepackage{ae,aecompl}

\usepackage{graphicx}	% Including figure files
\usepackage{amsmath}	% Advanced maths commands
\usepackage{amssymb}	% Extra maths symbols
\usepackage{booktabs}
\usepackage{multirow}
\usepackage{hyperref}
\usepackage{ulem}
\usepackage[super]{nth}
\newcommand{\gadget}[1]{\textsc{Gadget-2}#1} % typeset for Gadget2
\newcommand{\galic}[1]{\textsc{GalIC}#1} % typeset for GalIC
\newcommand{\simspin}[1]{\textsc{SimSpin}#1} % typeset for SimSpin

\newcommand{\eagle}[1]{\textsc{Eagle}#1}
\newcommand{\prospect}[1]{\textsc{ProSpect}#1}
\newcommand{\vsigma}[1]{$V/\sigma$#1}
\newcommand{\lR}[1]{$\lambda_{R}$#1}
\newcommand{\Reff}[1]{R$_{\text{eff}}$#1}
\newcommand{\psf}[1]{$\sigma_{\text{PSF}}$/R$_{\text{maj}}$#1}
\newcommand{\hyperfit}[1]{\textsc{hyper.fit}#1}

\newcommand{\sersic}[1]{S\'ersic{#1}}
\newcommand{\dlr}[1]{$\Delta \lambda_{R}${#1}}
\newcommand{\dvsig}[1]{$\Delta V/\sigma${#1}}
\newcommand{\obsdlr}[1]{$\Delta \lambda_{R}^{\text{obs}}${#1}}
\newcommand{\corrdlr}[1]{$\Delta \lambda_{R}^{\text{corr}}${#1}}
\newcommand{\obsdvsig}[1]{$\Delta V/\sigma^{\text{obs}}${#1}}
\newcommand{\corrdvsig}[1]{$\Delta V/\sigma^{\text{corr}}${#1}}

\title[Recovering $\lambda_R$ and $V/\sigma$]{Recovering $\lambda_R$ and $V/\sigma$ from seeing-dominated IFS data}

\author[K.E. Harborne et al.]{K.E. Harborne,$^{1,2}$\thanks{E-mail: katherine.harborne@icrar.org}
J. van de Sande,$^{2,3}$
L. Cortese,$^{1,2}$ 
C. Power,$^{1,2}$ \newauthor
A.S.G. Robotham,$^{1,2}$
C.D.P. Lagos,$^{1,2}$
S. Croom $^{2,3}$\\
$^{1}$International Centre for Radio Astronomy (ICRAR), M468, The University of Western Australia, 35 Stirling Highway, \\ 
Crawley, WA 6009, Australia\\
$^{2}$ARC Centre of Excellence for All Sky Astrophysics in 3 Dimensions (ASTRO 3D) \\
$^{3}$Sydney Institute for Astronomy, School of Physics, A28, The
University of Sydney, NSW, 2006, Australia \\
}

\date{Accepted XXX. Received YYY; in original form ZZZ}

\pubyear{2015}

\begin{document}
\label{firstpage}
\pagerange{\pageref{firstpage}--\pageref{lastpage}}
\maketitle

\begin{abstract}
Observers experience a series of limitations when measuring galaxy kinematics, such as variable seeing conditions and aperture size. These effects can be reduced using empirical corrections, but these equations are usually applicable within a restrictive set of boundary conditions (e.g. \sersic{} indices within a given range) which can lead to biases when trying to compare measurements made across a full kinematic survey. In this work, we present new corrections for two widely used kinematic parameters, \lR{} and \vsigma, that are applicable across a broad range of galaxy shapes, measurement radii and ellipticities. We take a series of mock observations of N-body galaxy models and use these to quantify the relationship between the observed kinematic parameters, structural properties and different seeing conditions. Derived corrections are then tested using the full catalogue of galaxies, including hydro-dynamic models from the \eagle{} simulation. Our correction is most effective for regularly-rotating systems, yet the kinematic parameters of all galaxies -- fast, slow and irregularly rotating systems -- are recovered successfully. We find that \lR{} is more easily corrected than \vsigma, with relative deviations of 0.02 and 0.06 dex respectively. The relationship between \lR{} and \vsigma, as described by the parameter $\kappa$, also has a minor dependence on seeing conditions. These corrections will be particularly useful for stellar kinematic measurements in current and future integral field spectroscopic (IFS) surveys of galaxies.
\end{abstract}

\begin{keywords}
galaxies: evolution -- galaxies: kinematics and dynamics 
\end{keywords}

%%%%%%%%%%%%%%%%%%%%%%%%%%%%%%%%%%%%%%%%%%%%%%%%%%

%%%%%%%%%%%%%%%%% BODY OF PAPER %%%%%%%%%%%%%%%%%%

\section{Introduction}
Stellar kinematics are a key component in unlocking the mysteries of galactic formation and evolution \citep{deZeeuw1991StructureGalaxies, Cappellari2016StructureSpectroscopy}. Prior to the millennium, morphology was often categorised by the light distribution alone. Using this approach, early-type elliptical systems appear smooth and structure-less, ``red and dead'' \citep{Binney1998GalacticAstronomy}. When kinematics are incorporated, this arm of Hubble's tuning fork segments into many more branches. Using long-slit spectroscopy, it was shown that elliptical galaxies have a slow rotational component \citep{Illingworth1977RotationGalaxies, Bertola1989Evidence4845, Binney1978OnGalaxies} and flattened ellipticals rotate more quickly \citep{Davies1983TheGalaxies}. Using integral field spectroscopy, the variety of different kinematic states only increased, from those with regular rotation at a various speeds to irregular systems with features like decoupled cores and embedded disks \citep{Emsellem2007TheGalaxies, Emsellem2011TheRotators, Cappellari2007TheKinematics}. 

\textsc{SAURON} \citep{Bacon2001TheSpectrograph, deZeeuw2002TheResults} and \textsc{Atlas}$^{\text{3D}}$ \citep{Cappellari2011TheCriteria} were the first two-dimensional, spatially-resolved, kinematic surveys to begin unravelling the kinematic morphology-density relationship, investigating this variety of kinematic structure and building a picture of how these structures have grown and evolved over time. They used two kinematic parameters to classify the kinematic-morphology of each system observed, \lR{} and \vsigma{}. Both quantities are used to understand the importance of random versus ordered motions in a galaxy. The observable spin parameter \lR{} was designed by \cite{Emsellem2007TheGalaxies} to better distinguish internal kinematic structure due to the radial dependence that takes full advantage of the 2D kinematic information.  This \lR{} parameter is defined, 
\begin{equation}
\label{eq:lambdaR}
\lambda_R = \frac{\sum_{i=1}^{n_p} F_i R_i |V_i|}{\sum_{i=1}^{n_p} F_i R_i \sqrt{V_i^2 + \sigma_i^2}}.
\end{equation}

The quantity \vsigma, which measures the relative importance of rotation to dispersion, can be described by the definition put forward by \cite{Binney2005RotationRevisited} and \cite{Cappellari2007TheKinematics}, 
\begin{equation}
\label{eq:vsigma}
    V/\sigma = \sqrt \frac{\sum_{i=1}^{n_p} F_i V_i^2}{\sum_{i=1}^{n_p} F_i \sigma_i^2},
\end{equation}

where $F_i$ is the observed flux, $R_i$ is the circularised radial position, $V_i$ is the line-of-sight (LOS) velocity and $\sigma_i$ is the LOS velocity dispersion, all quantified per image pixel, $i$, and summed across the total number of pixels, $n_p$, within some measurement radius. 

The spin parameter, \lR, is commonly used to divide galaxies into kinematic classes. Galaxies with low \lR{} and high \lR, as measured within the boundary containing half the total light (i.e. the half-light isophote), were labelled by \cite{Emsellem2007TheGalaxies, Emsellem2011TheRotators} as slow rotators (SRs) and fast rotators (FRs) respectively. \cite{Cappellari2016StructureSpectroscopy} re-formalised these divisions within the spin versus ellipticity plane, using the formula, 
\begin{equation}
    \lambda_{Re} \leq 0.08 + \varepsilon_{e}/4, \; \; \text{where} \; \varepsilon_{e} < 0.4;
    \label{eq:FRcriteria}
\end{equation}
where $\varepsilon_{e}$ is the ellipticity of the half-light isophote, \Reff, within which \lR{}$_{e}$ is calculated. SRs occupy the lower left hand corner of the \lR{}-$\varepsilon$ diagram with round, low ellipticities and often with irregular kinematic morphologies. The majority of galaxies appear as FRs, however, which occupy the rest of the parameter space. With these definitions, kinematic classes can be mapped out and trends between their distribution and other galaxy properties linked. 

Multi-object, integral field spectroscopy (IFS) surveys such as the \textsc{SAMI} survey \citep[Sydney-AAO Multi-object Integral field spectrograph;][]{Croom2012TheSpectrograph, Bryant2015TheSelection} and \textsc{MaNGA} \citep[Mapping Nearby Galaxies at Apache Point;][]{Bundy2015OverviewObservatory, Blanton2017SloanUniverse} are beginning to explore the nuances of kinematic morphology, with $\sim3,000$ ($z < 0.12$) and $\sim10,000$ ($z < 0.15$) galaxies respectively. These surveys have drawn links between the distribution of kinematic structures, stellar mass, local environment and age \citep{Emsellem2011TheRotators, Cappellari2011TheRelation, Cappellari2013TheFunction, Bois2011TheCores, Veale2017TheGalaxies, VanDeSande2018AShapes}.  These relationships have also been probed in cosmological simulations \citep{Jesseit2009Specificlambda_R-Parameter, Lagos2018TheGalaxies, vandeSande2019TheSimulations, Rosito2019TheGalaxies}. However, the dominant driver for transforming galaxies is still unclear; does the environment of a galaxy have any effect on the occurrence of different kinematic morphologies, or is galaxy mass a more important factor? Does the significance of these dependencies evolve across cosmic time? \citep{Penoyre2017TheSimulation, Brough2017TheClusters, Greene2018SDSS-IVGalaxies, Lagos2018QuantifyingGalaxies, Lagos2018TheGalaxies}. 

Future observing runs and surveys will build the census of galaxies we need to answer these questions. Kinematics will be measured out to larger and larger radii across broader redshift ranges with superb resolution. For example, the secondary \textsc{MaNGA} sample \citep{Wake2017TheConsiderations} will observe $\sim$3300 galaxies at z < 0.15 out to 2.5 Reff. With the next-generation of instruments, such as \textsc{Hector} \citep{Bryant2016Hector:Telescope}, the number of observations measured out to 2 Reff is set to increase dramatically; the \textsc{Magpi} survey\footnote{\url{http://magpisurvey.org/}} will observe $\sim$180 galaxies out at 0.25 < z < 0.35 out to 2-3 Reff using \textsc{MUSE}.

It has been demonstrated by a variety of groups, however, that our kinematic measurements are negatively affected by atmospheric seeing conditions \citep{DEugenio2013Fast0.183, vandeSande2017TheSurveys, vandeSande2017TheKinematics, Graham2018SDSS-IVProperties, Greene2018SDSS-IVGalaxies, Harborne2019Alambda_R}. The LOS velocity measurement is artificially decreased and LOS velocity dispersion increased due to beam smearing, causing measured values of \lR{} and \vsigma{} to decrease. When comparing measurements made at a variety of seeing conditions, as is often the case for surveys, it is unclear if the observed relationships are simply an artefact of observing conditions, or whether stronger trends would be observed if the measurements were corrected. 

In \cite{Graham2018SDSS-IVProperties} (hereafter G18), an empirical formula was presented that corrected measurements of \lR{} made within an effective radius, \Reff, for regular FRs. In \cite{Harborne2019Alambda_R}, it was demonstrated that this correction works well for an independent set of isolated $N$-body galaxies of a variety of morphologies. While the G18 correction is very successful for FR galaxies with \sersic{} indicies between $0.5 < n < 6.5$, it was not tested outside of this range. Furthermore, a similar correction is not available for \vsigma{} and conversion from one to the other is not trivial if the relationship between the two is also dependent on seeing \citep{Emsellem2007TheGalaxies, Cortese2019ThePhase}. 

Given the importance of the kinematic morphology-density relation in understanding galactic formation and evolution, it is important that we can apply corrections to all systems in a kinematic survey. The main goal of this paper is to design a seeing correction for \lR{} and \vsigma{} that are applicable across a broad range of galaxy shapes, measurement radii and projected inclination. Furthermore, we aim to test the accuracy of this correction and investigate whether possible systematic biases arise in the corrected sample. In Section \ref{sec:method}, we introduce our simulations and our methodology for observing these models. We present our fitting procedure and derived correction in \S \ref{sec:method-correct}. The results of applying this correction can be seen in \S \ref{sec:results} for fast, slow and irregular rotators. We also discuss the effect of seeing on the relationship between \lR{} and \vsigma. Overall conclusions can be found in \S \ref{sec:conclusions}. Throughout this work, we assume a Lambda-cold dark matter ($\Lambda CDM$) cosmology with $\Omega_m = 0.308, \Omega_{\Lambda} = 0.692$ and $H_0 = 67.8$.

\section{Method}
\label{sec:method}

Here we describe how we have constructed the data set used to derive and validate our corrections. This is divided into three parts: first, we explain the design of the galaxy catalogue; next, we outline how the full catalogue of galaxies has been constructed; finally, we describe how we have generated the synthetic IFS data-cubes, observed galaxy properties such as the effective radius (\Reff) and ellipticity ($\varepsilon$), and measured the kinematics \lR{} and \vsigma{} for all models. 

\bigskip{}

\subsection{Designing the catalogue}
\label{sec:catalogue}
The majority of galaxies in the Universe appear to be regular, FRs \citep{Graham2019SDSS-IVGroups/clusters, vandeSande2017TheSurveys, Cappellari2011TheRelation}. The SAMI survey contains $\sim 10\%$ SR following aperture correction \citep{vandeSande2017TheSurveys}, \textsc{Atlas}$^{\text{3D}}$ (selected for early type galaxies) contains $\sim 4 - 11\%$ \citep{Emsellem2011TheRotators}, and MaNGA contains $\sim 1 - 7\%$ \citep{Graham2019SDSS-IVGroups/clusters}. \textsc{Califa} \citep[Calar Alto Legacy Integral Field Area;][]{Sanchez2012CALIFAPresentation} contains 28\% SRs for stellar masses above $10^{11}$ M$_{\odot}$ \citep{Falcon-Barroso2019TheSequence}. SR fractions only become high in the most massive regimes around $10^{12}$ M$_{\odot}$ where the value was shown to go up to $\sim90\%$ by the \textsc{Massive} survey \citep{Veale2017TheGalaxies}. For this reason, it seems sensible to optimise our correction to work best for the regular FR class. 

We define SRs using the criteria from \cite{Cappellari2016StructureSpectroscopy}, as shown in Equation \ref{eq:FRcriteria}. In the alternative case that an observation is greater than this criterion for round isophotes ($\varepsilon_{e} < 0.4$), or flatter than $\varepsilon_{e} = 0.4$, the system is classed as an FR.\footnote{The boundary of $\varepsilon_{e} = 0.4$ is based on observations made by \textsc{Atlas}$^{\text{3D}}$ that all disk-less SR are rounder than $\varepsilon_{e} = 0.4$. This has been further supported by \textsc{Sami} \citep{Fogarty2015The2399} and \textsc{Califa} observations \citep{Falcon-Barroso2017StellarCorrections}.} For measurements of \lR{} made at greater or smaller radii than \Reff, the system still retains the classification made at \Reff.

Using N-body simulations, we can generate a wide variety of visual morphologies (i.e. E-S0 to Sd systems) with regularly-rotating velocity structures. These systems sit in equilibrium and describe the ``perfect'' isolated regular-rotator case. We have generated a sample of 18 models, shown in Table \ref{tab:sims}, spanning the visual morphology parameter space. Of this sample, we expect the S0-Sd galaxies to sit within the FR regime. The three E-S0 galaxies sit closer to the SR/FR division. 

We aim to apply our corrections to the full range of galaxies observed in a survey. This will include systems which have irregular kinematic morphologies, such as 2-$\sigma$ galaxies (where two dispersion maxima are seen near the centre of the galaxy in the two-dimensional LOS velocity and dispersion maps). Similarly, real galaxies in the universe may not be fully relaxed, equilibrium structures with regular velocity fields. 

To validate our corrections for the variety of kinematic classes, we have selected a further seven galaxies from the cosmological, hydro-dynamical simulation, \eagle{} \citep{Schaye2015TheEnvironments, Crain2015TheVariations, McAlpine2016TheCatalogues}. These galaxies are shown in Table \ref{tab:eagle}. While they have been selected because they are reasonably isolated systems at the present day, these systems have grown from cosmological initial conditions via mergers and accretions, as well as experiencing interactions. This provides us with a complementary sample of simulated galaxies whose kinematics are shaped by cosmologically realistic assembly histories.  

This gives us a full catalogue of 25 model galaxies. Because we have the full three-dimensional model, we can rotate and project each of these systems at a variety of different angles and measure the kinematics within various apertures. Each galaxy is observed multiple times in order to build up a comprehensive picture of the kinematic parameter space. 

\begin{table*}
\centering
\caption{Outlining the properties of each idealised isolated galaxy model created for use in this investigation. All models in this Table have been constructed using \galic{} and evolved for 10 Gyr using \gadget, are composed of 2.5$\times10^{6}$ particles and have a total stellar mass of $1 \times 10^{10}$ M$_{\odot}$. Images of all of these systems are shown in the supplementary materials in Appendix \ref{app:ch5_catalogue}.}
\begin{tabular}{@{}cccccc@{}}
\toprule
\textbf{Model} & \textbf{B/T} & \textbf{\begin{tabular}[c]{@{}c@{}}Concen-\\ tration,\\ $c$\end{tabular}} & \textbf{\begin{tabular}[c]{@{}c@{}}Bulge Scale \\ Length, \\ $A$ (kpc)\end{tabular}} & \textbf{\begin{tabular}[c]{@{}c@{}}Disk Scale \\ Length, \\ $h$ (kpc)\end{tabular}} & \textbf{\begin{tabular}[c]{@{}c@{}}S\'ersic Index, \\ $n$\end{tabular}} \\ \midrule
 &  & 1 & 62.149 & 9.279 & 5.01 \\
E-S0 & 0.8 & 1 & 31.075 & 9.279 & 6.42\\
 &  & 1 & 12.430 & 9.279 & 7.99 \\ \midrule
 &  & 1 & 12.430 & 9.279 & 2.81 \\
S0 & 0.6 & 10 & 3.451 & 5.640 & 3.97 \\
 &  & 50 & 0.972 & 2.914 & 5.36 \\ \midrule
 &  & 1 & 12.430 & 9.279 & 2.16 \\
Sa & 0.4 & 10 & 3.451 & 5.640 & 2.99 \\
 &  & 50 & 0.972 & 2.914 & 4.07 \\ \midrule 
 &  & 1 & 12.430 & 9.279 & 1.75 \\
Sb & 0.25 & 10 & 3.451 & 5.640 & 2.28 \\
 &  & 50 & 0.972 & 2.914 & 2.94 \\ \midrule
 &  & 1 & 12.430 & 9.279 & 1.24 \\
Sc & 0.05 & 10 & 3.451 & 5.640 & 1.26 \\
 &  & 50 & 0.972 & 2.914 & 1.46 \\ \midrule
 &  & 1 & 12.430 & 9.279 & 1.19 \\
Sd & 0.025 & 10 & 3.451 & 5.640 & 1.11 \\
\multicolumn{1}{l}{} & \multicolumn{1}{l}{} & 50 & 0.972 & 2.914 & 1.22  \\ \bottomrule
\end{tabular}
\label{tab:sims}
\end{table*}

\subsection{The Simulations}
\label{sec:method-sims}
\subsubsection{Isolated N-body models}

The initial conditions for each of the 18 N-body galaxy models have been constructed using \galic{} \citep{Yurin2014AnEquilibrium} and evolved for 10 Gyr using a modified version of \gadget{} \citep{Springel2005ModellingMergers}. 

\galic{} uses elements of made-to-measure \citep{Syer1996Made-to-measureSystems, Dehnen2009TailoringMethod} and Schwarzschild's techniques \citep{Schwarzschild1979AEquilibrium} to construct a bound system of particles that satisfy a stationary solution to the collision-less Boltzmann equation. Each model is initialised with three components: a dark matter halo distribution, a stellar bulge and a stellar disk. The dark matter halo, $\rho_{dm}(r)$, and stellar bulge, $\rho_{b}(r)$, structures are described by a Hernquist profile:
\begin{align}
    \rho_{dm}(r) &= \frac{M_{dm}}{2\pi}\frac{a_{dm}}{r(r+a_{dm})^3}, \\
    \rho_{b}(r) &= \frac{M_{b}}{2\pi}\frac{a_{b}}{r(r+a_{b})^3},
\end{align} 

where, for each component $i$, $M_i$ describes the total mass, $a_i$ the scale radius (which is a function of the chosen concentration), and $r$ the spherical radius defined with respect to the centre of mass. Stellar disks are generated with exponential profiles and an axis-symmetric velocity structure:
\begin{equation}
    \rho_{d}(R, z) = \frac{M_d}{4\pi z_0h^2}\text{sech}^2\left(\frac{z}{z_0}\right)\text{exp}\left(-\frac{R}{h}\right),
\end{equation}

where $R$ is the radius within the plane of the disk, $z$ is the height off the plane, $h$ is the scale length and $z_0$ is the scale height. In each case, the stellar component of the galaxy model contains $2.5\times10^{6}$ particles each with a mass of $4 \times 10^{3}$ M$_{\odot}$, corresponding to a total stellar mass of $1 \times 10^{10}$ M$_{\odot}$. The proportion of mass in the bulge and disk is determined by the bulge-to-total mass ratio (B/T). We have examined a variety of different B/T and concentrations, as shown in Table \ref{tab:sims}. Disks retain a smooth structure, with no spiral arms or features forming at this mass. We associate each particle with a luminosity based on a mass-to-light ratio of $1 \Upsilon_{\odot}$. We tested the impact of varying mass-to-light ratio between the bulge and the disk using a wide range in M/L of the two components, but we did not detect a significant effect from the results presented in this work (i.e. $\sim$ 0.03 dex residuals on corrected kinematics in the most extreme case considered).

We ensure that these models are kinematically equilibrated and numerically stable by evolving the particle positions for 10 Gyr using a modified version of \gadget{}. We have removed the particles that describe the dark matter component from the simulation and replaced them with an analytic form of the underlying dark matter potential. This ensures that the disk is stable against numerical artefacts caused by mass differences between stellar and dark matter particles \citep{Ludlow2019EnergySizes}; it also allows us to generate relatively high resolution models of regularly-rotating systems at low computational cost. The validity of this method has been evaluated in \cite{Harborne2019Alambda_R} and we direct the reader to this paper for further discussion. 

\begin{table*}
\centering
\caption{Outlining the properties of the \eagle{} galaxies included in this work. All galaxies are taken from the publically available \texttt{RefL0100N1054} simulation run. We show images of all of these systems in the supplementary materials in Appendix \ref{app:ch5_catalogue}.}
\label{tab:eagle}
\begin{tabular}{ccccccc}
\hline
\textbf{Galaxy ID} & \textbf{\begin{tabular}[c]{@{}c@{}}Group/ \\ Subgroup \\ Number\end{tabular}} & \textbf{\begin{tabular}[c]{@{}c@{}}Stellar \\ R$_{1/2}$, kpc\end{tabular}} & \textbf{\begin{tabular}[c]{@{}c@{}}S\'ersic \\ Index\end{tabular}} & \textbf{\begin{tabular}[c]{@{}c@{}}Stellar Mass,\\ $10^{10}$ M$_{\odot}$\end{tabular}} & \textbf{\begin{tabular}[c]{@{}c@{}}Stellar \\ Particle \\ Number\end{tabular}} & \textbf{\begin{tabular}[c]{@{}c@{}}Kinematic\\ Classification\end{tabular}} \\ \hline
9267523 & 1387/0 & 2.33 & 6.93 & 1.20 & 10028 & Low Rotation \\
10048611 & 1883/0 & 3.72 & 15.10 & 1.41 & 10994 & Odd \\
10770392 & 2461/0 & 2.33 & 7.48 & 1.63 & 12789 & 2-$\sigma$ \\
14202037 & 141/0 & 5.78 & 4.69 & 15.6 & 112382 & FR \\
17199679 & 30/1 & 3.61 & 4.78 & 14.3 & 101484 & FR \\
18223768 & 119/1 & 4.01 & 2.49 & 14.5 & 107591 & FR \\
18294880 & 946/0 & 3.27 & 7.78 & 2.83 & 23948 & Prolate \\ \hline
\end{tabular}
\end{table*}

\subsubsection{EAGLE hydro-dynamical models}
\eagle{} is a suite of cosmological hydro-dynamical simulations designed to investigate the formation and evolution of galaxies. We have used seven galaxies from the publicly available \texttt{RefL0100N1504} simulation run; this simulation box is a cubic volume with a side length of 100 co-moving Mpc, with initial baryonic particle masses of $m_g = 1.81 \times 10^6$ M$_{\odot}$, and maximum gravitational softening lengths of $\varepsilon_{\text{prop}} = 0.70$ physical kpc. 

Of the galaxies chosen, the three FRs are from a selection of systems with high spin parameters (\lR{} > 0.6) as identified by \cite{Lagos2018TheGalaxies}. The remaining galaxies, \texttt{GalaxyIDs = 10770392} (2-$\sigma$), \texttt{10048611} (odd), \texttt{9267523} (low rotation) and \texttt{18294880} (prolate), have been selected for analysis from a subsequent database of 217 galaxies identified within the \texttt{RefL0100N1504} box for having irregular or SR kinematic morphologies (Lagos et al., in prep). All of these galaxies have a stellar mass above $1 \times 10^{10}$ M$_{\odot}$, and contain at least 10,000 particles to describe this stellar component. This conservative limit has been selected to ensure that the numerical noise in these systems is low; this limit is higher than the convergence limit found for \eagle{} systems in \cite{Lagos2017AngularEAGLE}. We examine the mock $r$-band cutout images to ensure systems are isolated and not interacting with other systems at $z=0$.

In \eagle{}, each stellar particle is initialized with a stellar mass described by the \cite{Chabrier2003GalacticFunction} initial mass function; metallicities are inherited from the parent gas particle and ages are recorded from formation to current snapshot time. To convert these stellar properties into an observed flux, we follow the method outlined in \cite{Trayford2015ColoursSimulation}. Using the GALEXEV synthesis models \citep{Bruzual2003Stellar2003} for simple stellar populations, we generate a spectral energy distribution (SED) and associated flux for each stellar particle. For this sample, we have used \prospect{} \citep{Robotham2020ProSpect:Histories} to generate SEDs by logarithmic-ally interpolating the GALEXEV models (which provide a discrete set of ages and metallicities, ranging from t = $10^5$ - $2\times10^{10}$ yr and from $Z_{*} = 10^{-4} - 0.05$, respectively). We find that, of the galaxies chosen,  1-10\% of the metallicities lie outside of the extremities of the BC03 range, and so we extrapolate in these cases, as in \cite{Trayford2015ColoursSimulation}.  

Beyond this point, we follow the same process for both the isolated N-body galaxies and the hydro-dynamical \eagle{} sample. 

\subsection{The observations}
\label{sec:method-obs}

We have taken a series of mock observations of the 25 galaxies shown in Table \ref{tab:sims} and Table \ref{tab:eagle} using the R-package \simspin{} \citep{Harborne2020SimSpinCubes}. This code takes N-body/SPH models of galaxies and constructs kinematic data-cubes like those produced in an IFS observation. This code is registered with the Astrophysics Source Code Library \citep{Harborne2019SimSpin:Simulations} and can be downloaded directly from GitHub\footnote{\url{https://github.com/kateharborne/SimSpin}}. Using this framework, we have explored a large parameter space that includes kinematic measurements made at a variety of seeing conditions, projected inclinations and measurement radii.  

\subsubsection{Quantifying observational properties}
\label{sec:method-obs-quant}

Initially, we need to define the effective radius, \sersic{} index and ellipticity for each galaxy in the catalogue. Observationally, this would be done using ancillary data from larger optical surveys rather than from the kinematic cubes produced with an IFS. Hence, for the N-body models, we make a series of high resolution flux maps in which we place each galaxy at a sufficient distance that the aperture size encompasses the entire face-on projected galaxy. Pixels are set to 0.05'' equivalent to the resolving power of the Hubble space telescope. We have done the same for the \eagle{} galaxies, but given the particle resolution, we instead mimic the resolution of KiDS images with pixels set to 0.2''. We then use \textsc{ProFit} \citep{Robotham2017ProFit:Images} to divide each galaxy image into a series of iso-photal ellipses by rank ordering the pixels and segmenting these into equally-spaced flux quantiles. 

We use the surface brightness profile and the isophotal ellipses produced by \textsc{ProFit} to measure the effective radius and determine the ellipticity of the region. The effective radius, \Reff, is taken to be the outer semi-major axis of the elliptical isophote containing 50\% of the total flux. Taking all pixels interior to this radius, we compute the ellipticity by diagonalising the inertia tensor to give the axial ratio, $q$, where ellipticity is defined, $\varepsilon = (1-q)$.

We measure \lR{} and \vsigma{} at a variety of measurement radii and so compute the ellipticity at incremental factors of \Reff{} (R$_{\text{eff}}^{\text{fac}}$ values including 0.5, 1, 1.5, 2 and 2.5). Across this broad range of radii, the ellipticity of the isophotes does change and so the ellipticity of our measurement area must also vary. Following the method above, we take all pixels contained within the outer radius of an isophote containing incrementally larger portions of the total flux (at 11 divisions from 25\% to 75\%) and compute $q$ as a function of radius. Because this is discretised by the flux portions examined, to determine the axial ratio at specific radii we fit a polynomial to the radial $q$-distribution and interpolate to predict ellipticity at each position. 

\subsubsection{Measuring observable kinematics}
\label{sec:FR/SR_class}

We have make two sets of kinematic measurements throughout this experiment: \lR{} and \vsigma{} (Equations \ref{eq:lambdaR} and \ref{eq:vsigma}). With the cubes output from \simspin, we calculate these kinematic parameters for each observation. 

We generate IFS data cubes at the resolution of the SAMI. Stellar kinematics have been measured in this survey using both the blue and red spectra, but because most absorption features are present at blue wavelengths, we use these specifications for creating our mock IFS cubes. SAMI has a 580V grating mounted on the blue arm of the AAOmega, giving a resolution of R $\sim1800$ \citep{Scott2018TheProducts} and covering a wavelength range of $3700-5700$\AA. Kinematic cubes have a spatial pixel size of 0.5'' and a velocity pixel size of 1.04\AA{} \citep{Green2018TheProducts}. The line spread function for spectra extracted from the blue arm of the spectrograph is well approximated by a Gaussian with full-width half-maximum (FWHM) of $2.65$\AA{} \citep{vandeSande2017TheKinematics}.

\begin{figure}
    \centering
    \includegraphics[width=0.99\columnwidth]{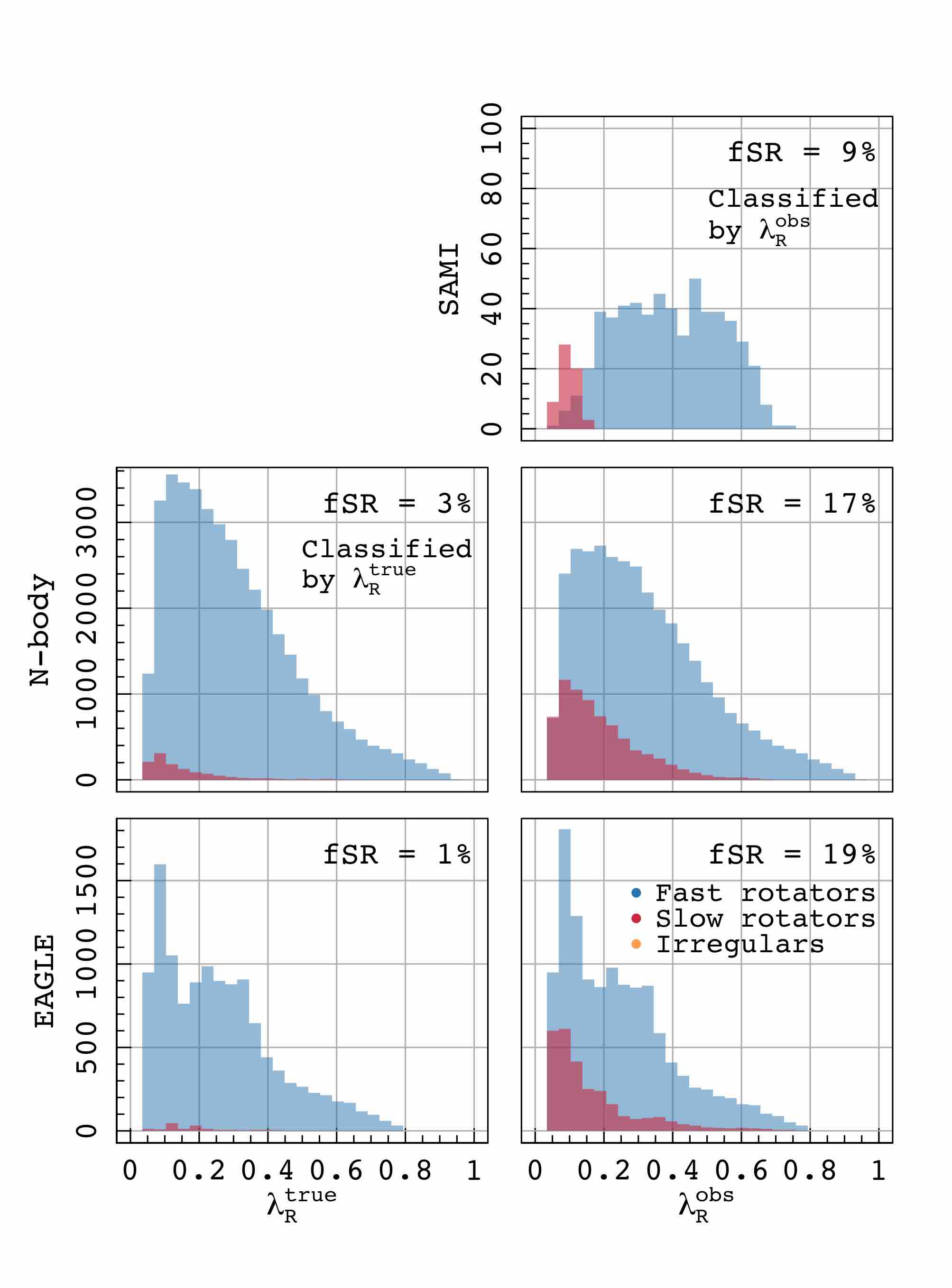}
    \caption{Demonstrating the distribution of \lR{} observations and their FR/SR classification made for the simulations in Table \ref{tab:sims} (centre) and Table \ref{tab:eagle} (below). On the left, we show the all observations with classification based on their ``true'' classification; on the right, we show classifications set by the observed \lR. FRs are shown in blue, SRs as red and irregular galaxies as orange. Each galaxy observation is classified by the equivalent observation at \Reff{} \citep{Cappellari2016StructureSpectroscopy}, and hence in the cases where kinematics are measured at R$_{\text{eff}}^{\text{fac}}$ = 2.5, \lR{} values may be quite high. The percentage of SR for the total distribution in each case is shown as ``fSR'' in the upper right corner. A comparison to the SAMI galaxies is made (above), where we show the distribution of measurements within SAMI DR2 \citep[using the quality criteria and cuts made in][]{vandeSande2017TheKinematics, vandeSande2019TheSimulations} and where the fSR fraction has been calculated as for the \simspin{} galaxies.}
    \label{fig:observations}
\end{figure}

For each galaxy, we generate images at 19 equally-spaced inclinations from $0 - 90^{\circ}$ (i.e. from face-on to edge-on, respectively); at each inclination, we have applied 31 equally-spaced degrees of seeing, increasing the FWHM of the Gaussian PSF from $0 - 15''$ (where $\sigma_{\text{PSF}} = \text{FWHM}_{\text{PSF}} / 2.355$). At each level of blurring, the measurement ellipse is held constant, as measured from our high resolution images explained in Section \ref{sec:method-obs-quant}. Finally, we have considered a range of measurement radii, taking our kinematic measurements of  \lR{} and \vsigma{} within 5 factors of the effective radius from $0.5 - 2.5$ \Reff. For each of these radius factors, the ellipticity of the corresponding isophote at this new distance has been modified and the kinematic value calculated within this new ellipse. 

Throughout these observations, we keep the spatial sampling within the measurement ellipse consistent. The spatial sampling and aperture size have a strong impact on the measurement of kinematics, as shown by  \cite{DEugenio2013Fast0.183} and \cite{vandeSande2017TheSurveys}. To make sure that our values are comparable (and that any measured differences are not due to the effects of spatial sampling), we have projected each galaxy at a distance such that the semi-major axis of the measurement ellipse is equivalent to the same number of pixels (e.g. 14 px within the 15px aperture radius). \footnote{For completeness, we have explored the spatial sampling dependence of our correction in Appendix \ref{app:ch5_spaxels} and demonstrate its validity and the corrected-kinematic uncertainties for a range of spatial sampling scenarios.}

This gives us 2945 measurements of \lR{} and \vsigma{} for each galaxy: 73,625 observations of each kinematic measure in total. The distribution of \lR{} measured in each of these is shown in Figure \ref{fig:observations}, in comparison to the distribution of kinematics from the SAMI DR2 \citep{Scott2018TheProducts}. 

We have classified these observations individually as FRs or SRs based on the criterion in Equation \ref{eq:FRcriteria}. Because \cite{Emsellem2007TheGalaxies} and \cite{Cappellari2016StructureSpectroscopy} defined this equation based on the measured kinematics within 1 \Reff, we have used the classifications from the R$_{\text{eff}}^{\text{fac}} = 1$ sample to label the observations made at other apertures (i.e. the kinematic class is defined using the \lR{} measurement at R$_{\text{eff}}^{\text{fac}} = 1$ but assigned to all measurement radii for a specific galaxy at a given inclination).

On the left, we show each observation with the labeled kinematic class based on the true kinematics; on the right, we show all observations with kinematic class based on the observed kinematics. This highlights one of the main concerns of seeing on the measurement of kinematics. Initially, when classifying observations made of the N-body systems at perfect seeing, 3\% are classified as SR. However, following the addition of atmospheric blurring, this percentage increases to 17\%. Observations of inherently FR systems are pushed into the SR category due to the additional dispersion of the atmosphere. Fractions such as these form the basis of many works on kinematic morphology-density evolution \citep{DEugenio2013Fast0.183, Houghton2013FastCluster, Scott2014DistributionCluster, vandeSande2017TheSurveys, Lagos2018TheGalaxies, Graham2019SDSS-IVGroups/clusters}, and this confirms the need for the seeing correction derived in this work. Note that the $N$-body catalogue was not designed to reproduce realistic kinematic-morphology. We show how this fraction changes from true to observed and finally following correction in order to justify the application of this work to real data.

Having generated 73,625 measurements of both \lR{} and \vsigma{} in a variety of observational conditions, we can now map the parameter space with \psf{}, fit a model in order to correct for these effects and then test the correction on a wide variety of different galaxy models and projection conditions. 

\begin{figure*}
    \centering
    \includegraphics[width=\textwidth]{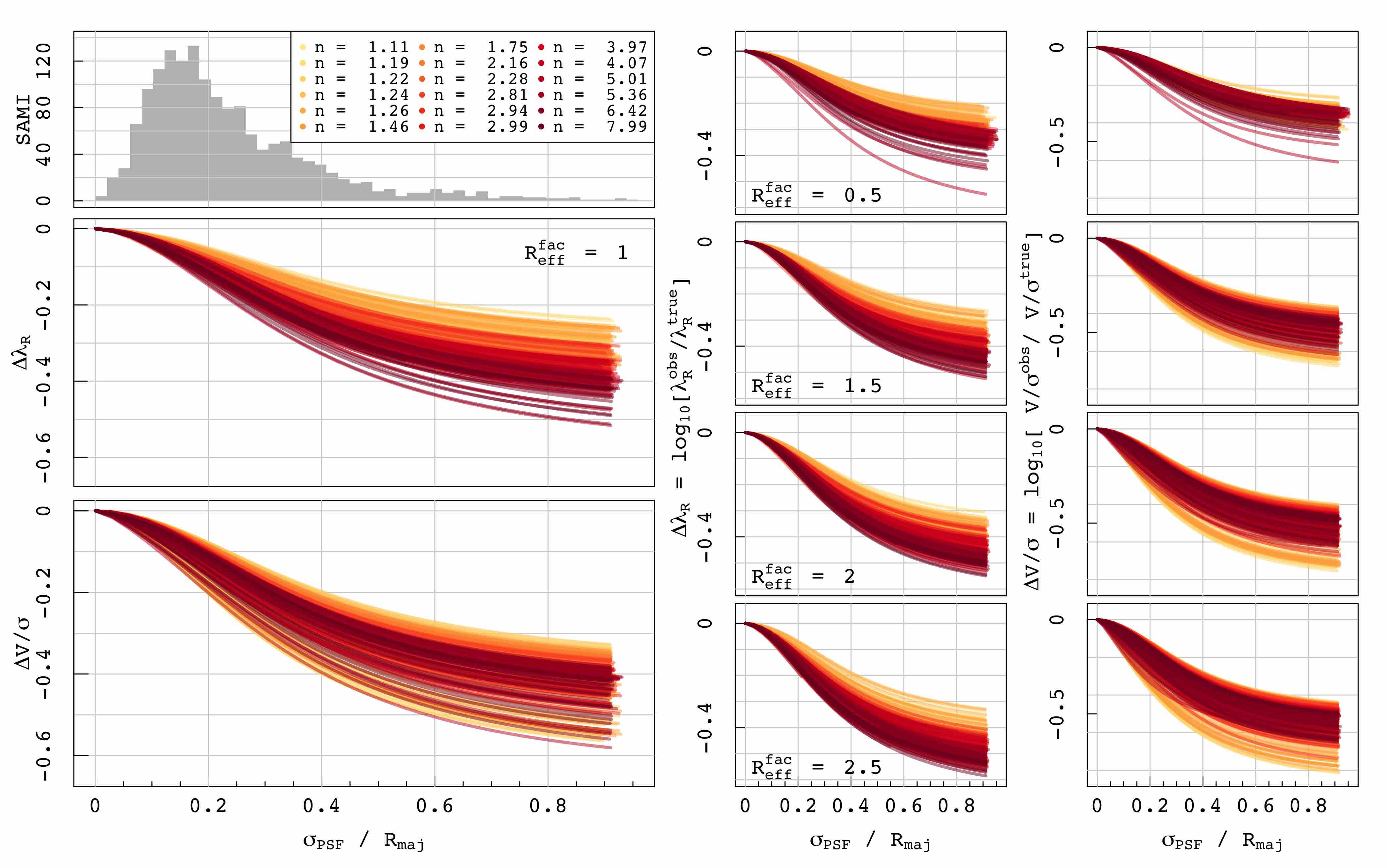}
    \caption{Demonstrating a selection of regular rotator kinematic tracks with increasingly poor seeing conditions as compared to the distribution of observations in the \textsc{SAMI} survey DR2. Colour indicates galaxy shape according to \sersic{} index. Here we can see that the relationship is well described by a power law, but the spread in this power law has strong dependencies on galaxy projection and shape. Similar trends are seen for $\text{R}_{\text{eff}}^{\text{fac}} = 0.5, 1.5, 2,$ and $2.5$, as shown on the right.}
    \label{fig:psf_dlr_dvsig}
\end{figure*}

\section{Modelling the correction}
\label{sec:method-correct}

\begin{figure*}
	\includegraphics[width=\textwidth]{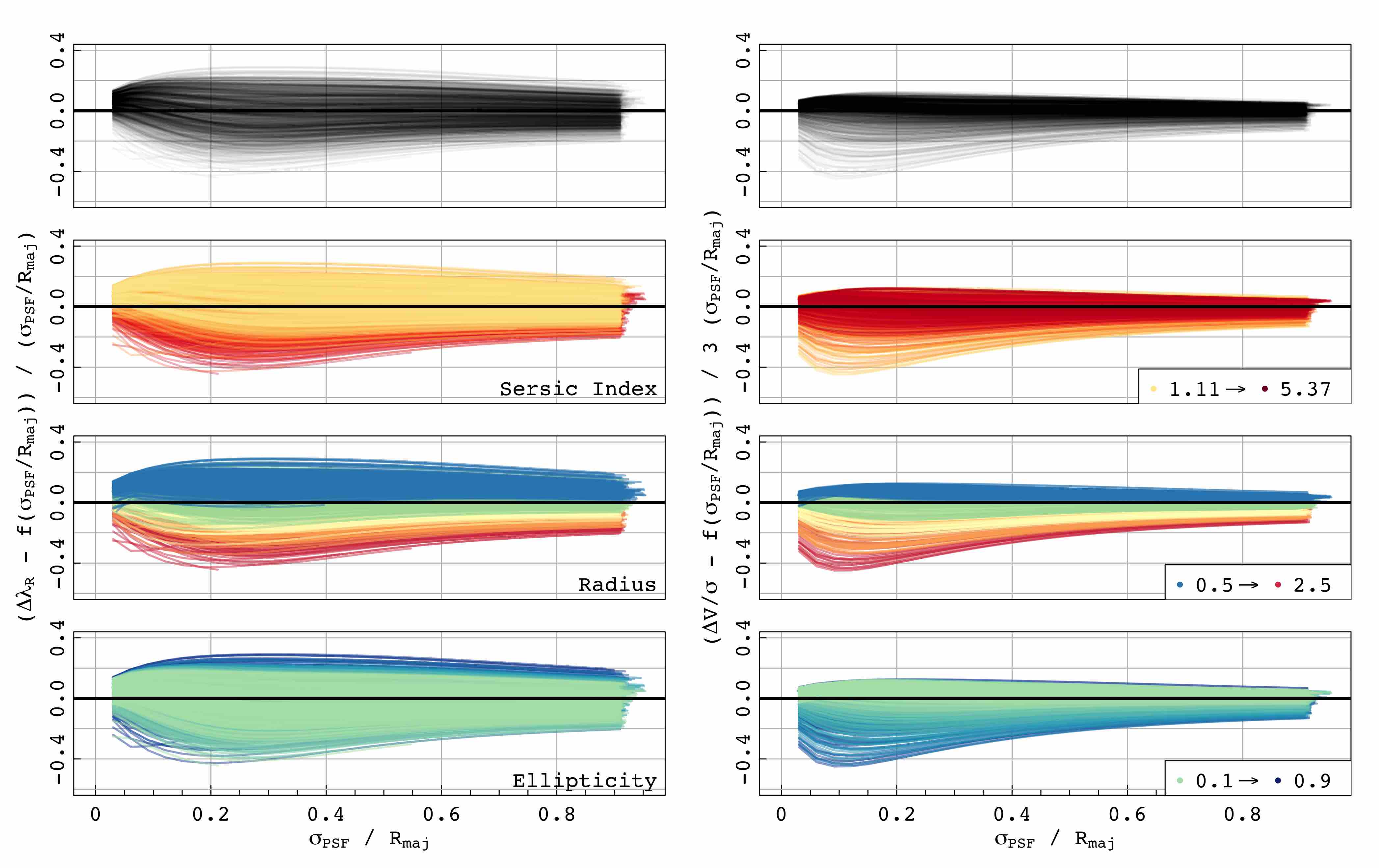}
    \caption{(Top row) Demonstrating the effect of subtracting the \psf{} dependence from the tracks for $\Delta \lambda_{R}$ (left) and \dvsig{} (right). (Lower three rows) Examining how residual tracks are influenced by observational effects of ellipticity, \sersic{} index and measurement radius. The predominant source of scatter appears to differ for \lR{} and \vsigma; the former seems strongly correlated with \sersic{} index and the latter more dependent on ellipticity.}
    \label{fig:sigmoid_fit_residuals}
\end{figure*}
From our catalogue of 73,625 observations, we only use the S0-Sd models from Table \ref{tab:sims} to model the correction. This allows us to design a correction that works best for the majority of galaxies, as discussed in Section \ref{sec:catalogue}. Hence, We focus on the isolated regular rotators that fall well within the FR regime, which reduces our sample to 44,175 observations and leaves the rest of the catalogue for verification.  

First, we define the ``true'' value for each kinematic measure. As in G18, the common assumption is that the intrinsic ``true'' \lR{} value corresponds to a measurement made in perfect seeing conditions. In this work, we extend this definition: \textbf{\lR$^{\text{true}}$ and \vsigma$^{\text{true}}$ are defined to be the value measured within a fixed measurement radius at a fixed inclination when seeing conditions are perfect.} We parameterise the seeing conditions by the ratio of the semi-major axis of the measurement ellipse, R$_{\text{maj}}$ relative to the $\sigma$ of the PSF (i.e. for perfect seeing, \psf{} = 0). 

Hence, the relative difference between the observed and true values is defined:
\begin{align}
\Delta \lambda_{R} &= \text{log}_{10}(\lambda_{R}^{\text{obs}}) - \text{log}_{10}(\lambda_{R}^{\text{true}}) \;, \label{eq:lr_true}\\ 
\Delta V/\sigma &= \text{log}_{10}(V/\sigma^{\text{obs}}) - \text{log}_{10}(V/\sigma^{\text{true}}) \;. \label{eq:vsig_true}
\end{align}

We describe our parameter space using these definitions, considering how \dlr{} and \dvsig{} change with \psf. Because inclination is difficult to parameterise in observations without modelling \citep[i.e.][]{Taranu2017Self-consistentKinematics}, we use the observed ellipticity of the galaxy at each inclination as a proxy. To mathematically describe the behaviour of kinematics with seeing, we make the following assumptions:
\begin{enumerate}
    \item That the parameter space for regular rotators (\psf{} versus \dlr{} or \dvsig{}) can be fully described using galaxy shape, as defined by \sersic{} index and ellipticity, and measurement radius.
    \item In the case where a galaxy is unresolved (i.e. when \psf{} $>$ 1), no attempts would be made to correct kinematic measurements.
    \item Equally, when seeing conditions are perfect (i.e. \psf{} = 0), no correction would be applied.
    \item We cannot reliably quantify the kinematics of face-on galaxies. Furthermore, kinematically there is a degeneracy between a face-on FR and an elliptical SR at any inclination. In this sample, we have removed tracks for observations where \lR{} or \vsigma{} are less than 0.05, and in cases were the observed ellipticity, $\varepsilon$ is less than 0.05 in order to account for this.
    \item Properties such as ellipticity and effective radius have been calculated accurately from high resolution data and, if required, corrected independently. We do not address the effects of observation on the recovery of these properties. See works by \cite{Cortese2016TheMorphology, Weijmans2014TheGalaxies, Jesseit2009Specificlambda_R-Parameter, Padilla2008TheSurvey, Krajnovic2006Kinemetry:Distribution}; etc. for further discussion of this topic.  
\end{enumerate}

\begin{figure}
	\includegraphics[width=\columnwidth]{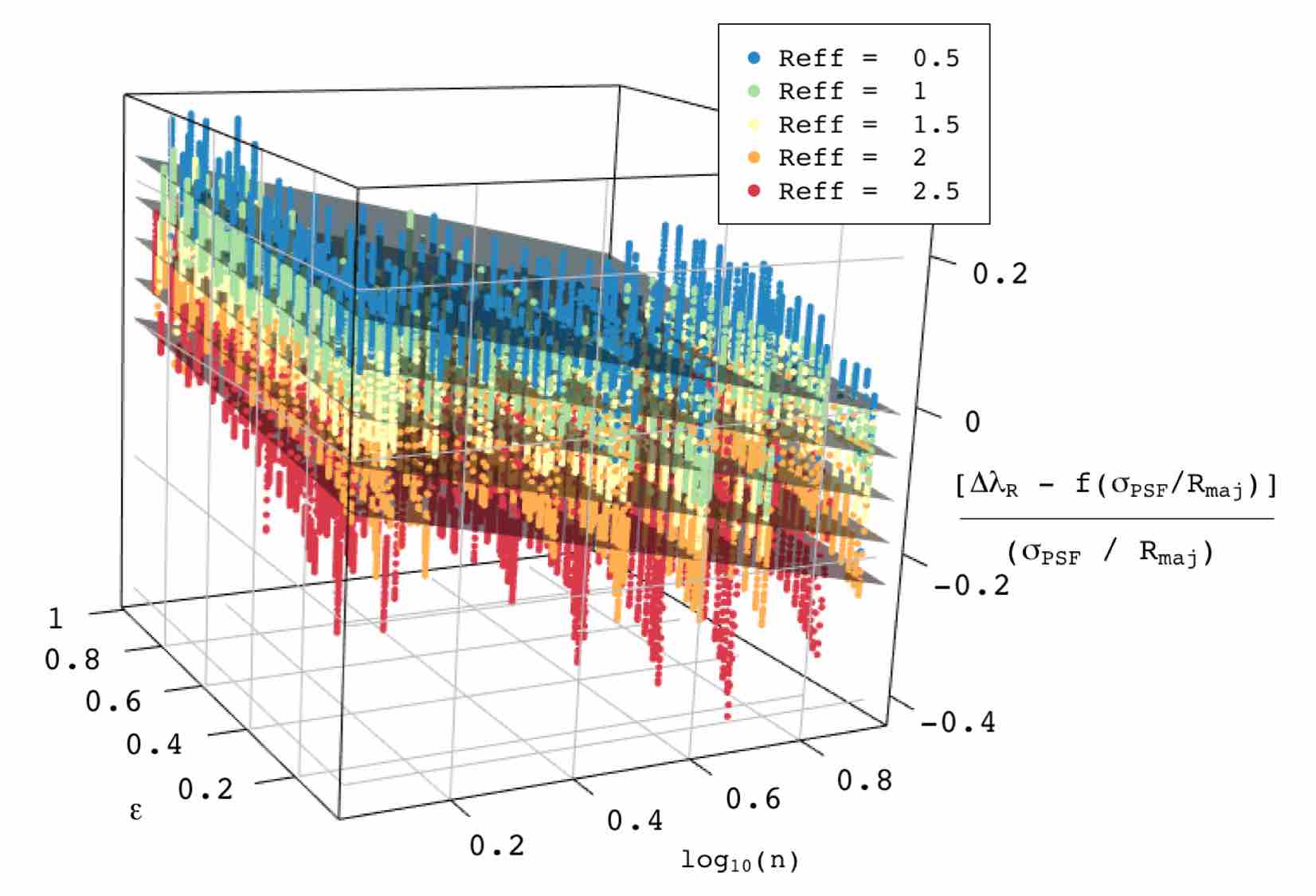}
    \caption{Demonstrating the \hyperfit{} plane fitted to the scatter in the residual of \dlr. As this is a four-dimensional data set, we show the scatter in 3D with colour representing the final parameter. In this case, each colour represents each R$_{\text{eff}}^{\text{fac}}$. When presented in this fashion, it is clearly appropriate to fit a hyper-plane to describe the data.}
    \label{fig:4D}
\end{figure}

\begin{figure*}
	\includegraphics[width=\textwidth]{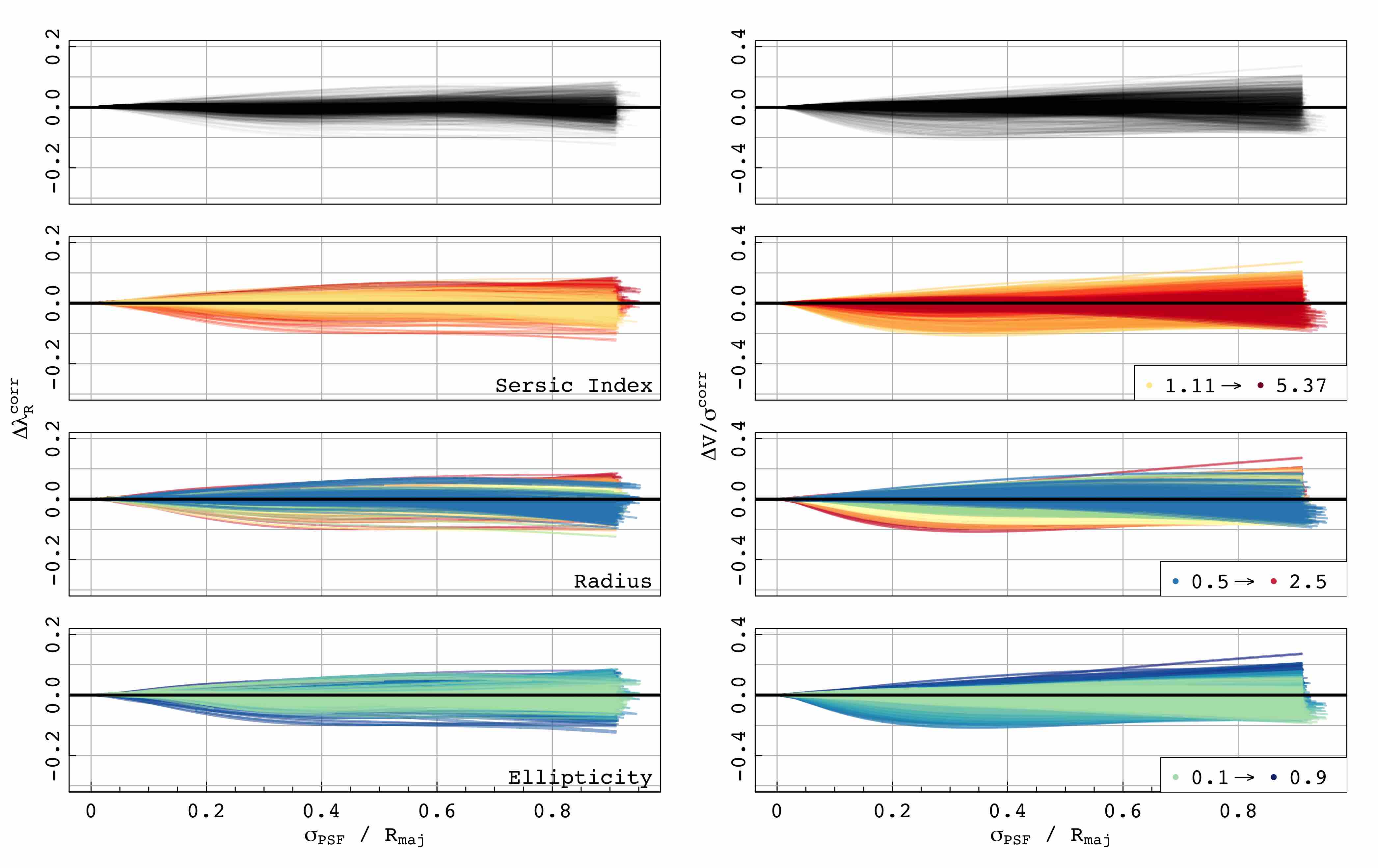}
    \caption{Examining how final residual tracks for \dlr{} (left) and \dvsig{} (right) are influenced by observational effects of ellipticity, \sersic{} index and measurement radius once the scatter has been accounted for with \hyperfit. We have greatly reduced the dependencies for this range of observational constraints ($\varepsilon, n, \text{R}_{\text{eff}}^{\text{fac}}$), though some minor scatter still remains.}
    \label{fig:sigmoid_fit_hf_residuals}
\end{figure*}

Following these assumptions, we are left with a sample of 35,499 observations of \lR{} and 35,567 observations of \vsigma{} from which we can derive corrections. We test the validity of the first assumption by examining the relationship between \psf{} and $\Delta$. Figure \ref{fig:psf_dlr_dvsig} shows a selection of \psf{} versus \dlr{} and \dvsig{} tracks for measurements made within 1 \Reff{} on the left hand side; very similar trends can be seen for the other $\text{R}_{\text{eff}}^{\text{fac}}$ values considered, as shown in the panels to the right. Each one of these tracks demonstrates how the kinematic measurement changes as seeing conditions worsen, combining the 31 observations at 0 $\leq$ \psf{} < 1 into a single track. 

Within both kinematic properties, we see that all tracks generally describe a classic ``S-curve'' sigmoid function. In Figure \ref{fig:psf_dlr_dvsig}, we have colour-coded tracks by their \sersic{} index. There is significant scatter in the exact parameterisation due to the shape, ellipticity and measurement radius of the galaxy being observed. Similar effects were observed in G18. The dependence on \sersic{} index appears to be inverted for \vsigma{} (in comparison to \lR). This is an interesting feature that may be due to the fact that \lR{} is dependent on the velocity measurement in both the numerator and the denominator of Equation \ref{eq:lambdaR}, effectively cancelling out some of the effects of seeing. We also see the gradient of each distribution scales with measurement radius. There may be further factors that contribute to the scatter in this plane, but these properties fairly represent the dominant sources of uncertainty that can be quantified observationally.

Hence, we can describe the behaviour of any track using two functions; one function that describes the sigmoid shape caused by seeing and a second that describes how different parameters influence the scatter in the residual, i.e. 
\begin{align}
    \Delta \lambda_{R} &\sim f\left(\frac{\sigma_{\text{PSF}}}{\text{R}_{\text{maj}}}\right) + f(\varepsilon, n, \text{R}_{\text{eff}}^{\text{fac}}), \label{eq:lr_pcor} \\
    \Delta V/\sigma &\sim f\left(\frac{\sigma_{\text{PSF}}}{\text{R}_{\text{maj}}}\right) + f(\varepsilon, n, \text{R}_{\text{eff}}^{\text{fac}}) \label{eq:vsig_pcor}.
\end{align}

First, we consider the sigmoid equation that describes the trend between the kinematics and \psf. We fit each track with a sigmoid function, given by:
\begin{align}
    f\left(\frac{\sigma_{\text{PSF}}}{\text{R}_{\text{maj}}}\right) &= \frac{\alpha}{1 + \text{exp}\left[\beta \left(\frac{\sigma_{\text{PSF}}}{\text{R}_{\text{maj}}}\right)^{\gamma} + \delta\right]} + c. \label{eq:sigmoid}
\end{align}

In order to facilitate our third assumption (that no correction is applied when seeing conditions are perfect), we set $c = -\alpha/(1+\text{exp}\left[ \delta \right])$. This provides additional constraints on the fit. By minimising the sum of square residuals, we optimise the fit of this function to each track in our sample (consisting of 1167 and 1166 tracks for \dlr{} and \dvsig{} respectively) and take the mean value of $\alpha$, $\beta$, $\gamma$, and $\delta$ in order to describe the average track shape. In doing so, we find the following best fitting models describe the general shape:
\begin{align}
    \begin{split}
    f\left(\frac{\sigma_{\text{PSF}}}{\text{R}_{\text{maj}}}\right)^{\Delta \lambda_{R}} &= \frac{7.48}{1 + \text{exp}[4.08 \left(\frac{\sigma_{\text{PSF}}}{\text{R}_{\text{maj}}}\right)^{1.60} + 2.89]} - 0.39 \label{eq:f_psf_lr} 
    \end{split} 
\end{align}
\begin{align}
    \begin{split}
    f\left(\frac{\sigma_{\text{PSF}}}{\text{R}_{\text{maj}}}\right)^{\Delta V/\sigma} &= \frac{7.55}{1 + \text{exp}[4.42 \left(\frac{\sigma_{\text{PSF}}}{\text{R}_{\text{maj}}}\right)^{1.55} + 2.73]} - 0.46 \label{eq:f_psf_vsig} 
    \end{split} 
\end{align}

Once the initial sigmoid has been subtracted, the residuals that remain in both \dlr{} and \dvsig{} are a flared distribution about zero with standard deviation of 0.05 and 0.08 dex respectively. Returning to our assumptions, we suggested that the scatter in the tracks is dependent on the observational parameters of galaxy shape, ellipticity and measurement radius.  In the lower panels of Figure \ref{fig:sigmoid_fit_residuals}, we investigate the remaining scatter, where the residuals have each been divided by some factor of the seeing conditions and coloured by the values used to parameterise galaxy shape. This divisor facilitates our assumption that the correction goes to zero at the appropriate bounds. In doing so, we find that the observational parameters appear much more linear, as the flare about zero has been removed, and can more efficiently be described using a hyper-plane.  

We use \hyperfit{} to fit the remaining scatter. This is an R-package developed by \cite{Robotham2015Hyper-Fit:Uncertainties} that recovers the best-fitting linear model by maximising the general likelihood function, assuming that some $N$-dimensional data set can be described by a ($N-1$)-dimensional plane with intrinsic scatter. Examining the data in Figure \ref{fig:sigmoid_fit_residuals}, and plotting the data in three dimensions, as in Figure \ref{fig:4D}, it seems an reasonable assumption that we can describe this distribution using a plane. 

\bigskip{}

There are a very large number of possible fitting routines contained within the \hyperfit{} package. Systematically, we checked all available algorithms and settled on the method that minimises the intrinsic scatter in the solution. In all cases, the hit-and-run (HAR) algorithm \citep{Garthwaite2010AdaptiveProcess} produced the lowest values of intrinsic scatter. 
\begin{align}
\begin{split}
    f(\varepsilon, n, \text{R}_{\text{eff}}^{\text{fac}})^{\Delta \lambda_{R}} = [0.10 \times \varepsilon] + [-0.22 \times \text{log}_{10}(n)] \\ + [-0.12 \times  \text{R}_{\text{eff}}^{\text{fac}}] + 0.22, \; (\sigma = 0.048), \label{eq:f_hf_lr} 
\end{split} \\
\begin{split}
    f(\varepsilon, n, \text{R}_{\text{eff}}^{\text{fac}})^{\Delta V/\sigma} = [-0.10 \times \varepsilon] + [0.024 \times \text{log}_{10}(n)] \\ + [-0.056 \times  \text{R}_{\text{eff}}^{\text{fac}}] + 0.12, \; (\sigma = 0.047). \label{eq:f_hf_vsig}
\end{split}
\end{align}

\begin{figure}
	\includegraphics[width=\columnwidth]{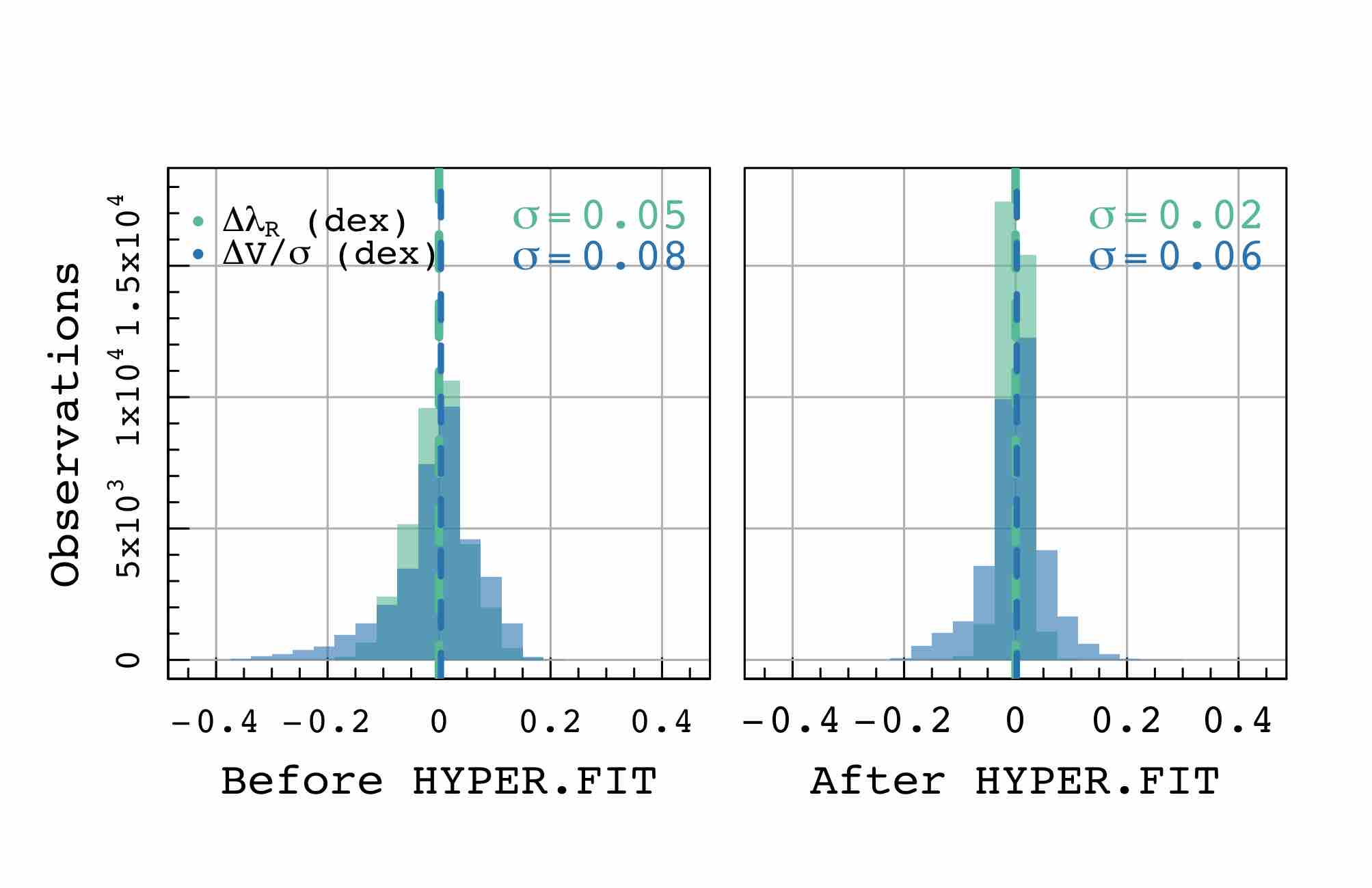}
    \caption{Demonstrating the relative residuals in \dlr{} and \dvsig{} before and after the \hyperfit{} correction is applied. When we take galaxy shape into account using \hyperfit{}, we find a more balanced distribution about zero. The residals for \lR{} are smaller as compared to \vsigma{}, which shows that we correct \lR{} with more success.}    
    \label{fig:rel_res}
\end{figure}

Subtracting these final trends from the residuals in Figure \ref{fig:sigmoid_fit_residuals}, we find that the dependencies with shape are majorly removed, as shown in Figure \ref{fig:sigmoid_fit_hf_residuals}. These distributions have mean of zero before and after correction, but the standard deviation for these corrected tracks is 0.02 and 0.06 dex (improved from $\sigma =$ 0.05 and 0.08) for \dlr{} and \dvsig{} respectively. We show a relative comparison of these distributions before and after the \hyperfit{} correction in Figure \ref{fig:rel_res}.

Substituting in these hyper-plane expressions into Equations \ref{eq:lr_pcor} and \ref{eq:vsig_pcor}, we have constructed the full corrections for \dlr{} and \dvsig{}. 

\begin{center}
    \fbox{\begin{minipage}{.9\columnwidth}
\begin{equation} 
    \Delta \lambda_{R}^{\text{corr}} = f\left(\frac{\sigma_{\text{PSF}}}{\text{R}_{\text{maj}}}\right) + \left(\frac{\sigma_{\text{PSF}}}{\text{R}_{\text{maj}}}\right) \times f(\varepsilon, n, \text{R}_{\text{eff}}^{\text{fac}}), \label{eq:lr_corr}
\end{equation}

\noindent where,
\begin{align}
    \begin{split}
    f\left(\frac{\sigma_{\text{PSF}}}{\text{R}_{\text{maj}}}\right)^{\Delta \lambda_{R}} = \frac{7.48}{1 + \text{exp}[4.08 \left(\frac{\sigma_{\text{PSF}}}{\text{R}_{\text{maj}}}\right)^{1.60} + 2.89]} - 0.39,
    \end{split} \nonumber 
\end{align} \\
\begin{align}
    \begin{split}
    f(\varepsilon, n, \text{R}_{\text{eff}}^{\text{fac}})^{\Delta \lambda_{R}} = [0.10 \times \varepsilon] + [-0.22 \times \text{log}_{10}(n)] \\ + [-0.12 \times  \text{R}_{\text{eff}}^{\text{fac}}] + 0.22. \nonumber
    \end{split}
\end{align}
\end{minipage}
}
\end{center}

\begin{center}
    \fbox{\begin{minipage}{.9\columnwidth}
\begin{equation}
    \Delta V/\sigma^{\text{corr}} = f\left(\frac{\sigma_{\text{PSF}}}{\text{R}_{\text{maj}}}\right) + 3 \left(\frac{\sigma_{\text{PSF}}}{\text{R}_{\text{maj}}}\right) \times f(\varepsilon, n, \text{R}_{\text{eff}}^{\text{fac}}), \label{eq:vsig_corr}
\end{equation}

\noindent where,
\begin{align}
    \begin{split}
    f\left(\frac{\sigma_{\text{PSF}}}{\text{R}_{\text{maj}}}\right)^{\Delta V/\sigma} = \frac{7.55}{1 + \text{exp}[4.42 \left(\frac{\sigma_{\text{PSF}}}{\text{R}_{\text{maj}}}\right)^{1.55} + 2.73]} - 0.46,
    \end{split} \nonumber \\
    \begin{split}
     f(\varepsilon, n, \text{R}_{\text{eff}}^{\text{fac}})^{\Delta V/\sigma} = [-0.10 \times \varepsilon] + [0.024 \times \text{log}_{10}(n)] \\ + [-0.056 \times  \text{R}_{\text{eff}}^{\text{fac}}] + 0.12. \nonumber
    \end{split}
\end{align}
\end{minipage}
}
\end{center}

In order to arrive at the $\lambda_R^{\text{corr}}$ and $V/\sigma^{\text{corr}}$ values, we also need to invert equations \ref{eq:lr_corr} and \ref{eq:vsig_corr}. This is shown in Equation \ref{eq:invert} for completeness. To ease the conversion of measured to corrected values, we also present a simple Python code for public use available on GitHub\footnote{\url{http://github.com/kateharborne/kinematic_corrections}}.

\begin{align}
    \lambda_R^{\text{corr}} &= 10^{\left[\text{log}_{10}(\lambda_R^{\text{obs}}) - \Delta \lambda_{R}^{\text{corr}}\right]}, \nonumber \\
    V/\sigma^{\text{corr}} &= 10^{\left[\text{log}_{10}(V/\sigma^{\text{obs}}) - \Delta V/\sigma^{\text{corr}}\right]}. \label{eq:invert}
\end{align}

\begin{figure*}
	\includegraphics[width=\textwidth]{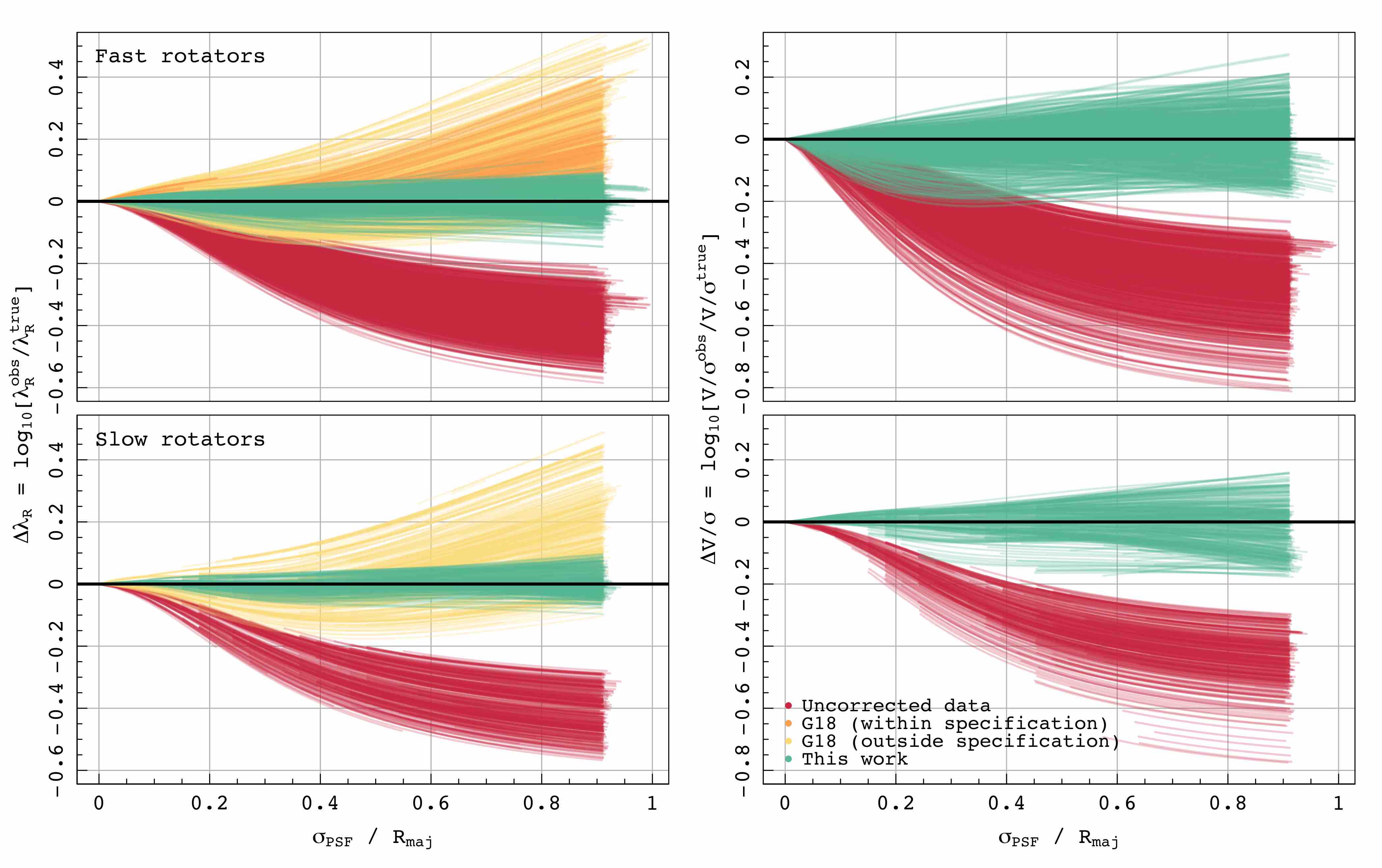}
    \caption{Demonstrating the success of the correction for regular FRs (above) and for SRs (below). The panels on the left show the observed \dlr{} values prior to correction (red), G18's empirical correction within the bounds described in G18 (orange), G18's correction but for data outside the bounds ($n > 6.5$ or R$_{\text{eff}} \neq 1$) (yellow), and the \hyperfit{} correction derived through this work (green). We show the same distribution but for \dvsig{} in the panels on the right.}    
    \label{fig:hf_correction}
\end{figure*}

\section{Results}
\label{sec:results}

Having derived these corrections using the 15 S0-Sd galaxies from our N-body catalogue, we test how effective our correction is for all galaxies in the catalogue. In this section, we begin by examining how effective this correction is on the full $N$-body catalogue, breaking these observations down into the FR and SR classes as given by Equation \ref{eq:FRcriteria}. As a separate test, we investigate how well the correction works for galaxies from the \eagle{} simulation, using divisions of FR, SR and a further class of irregular systems. Finally, we use our extensive data set to examine how the \lR{} and \vsigma{} kinematic parameters are related and whether this has any dependence on seeing conditions. 

\subsection{N-body catalogue results}
\label{sec:res_Nbody}

We begin with a sample of 53,010 observations of \lR{} and \vsigma{} for the 18 galaxies in Table \ref{tab:sims}. We remove any values for which \lR{} or \vsigma{} is less than 0.05, as well as any where the ellipticity is less than 0.05 (as explained in Section  \ref{sec:method-correct}). This leaves us with a total sample of 41,202 and 41,325 observations for \lR{} and \vsigma{} respectively, including the three E-S0 regular rotators. 

We divide our observations into FRs and SRs using the criteria of \cite{Cappellari2016StructureSpectroscopy}, as shown in Equation \ref{eq:FRcriteria}. As described in Section \ref{sec:FR/SR_class}, these classifications are based on the measurements of \lR{} made at R$_{\text{eff}}^{\text{fac}} = 1$. This gives a sample of 34,039 FR and 7,163 SR observations. 

Figure \ref{fig:hf_correction} demonstrates the effect of applying corrections to both \lR{} (left) and \vsigma{} (right) of the N-body models in Table \ref{tab:sims}. For FR observations (above), this gives 1317 individual tracks for both \lR{} and \vsigma. As shown on the left, not only does our correction remove the negative trend with seeing for uncorrected \lR{}, it also significantly reduces the scatter of the distribution. This is because we have included the observed ellipticity which effectively accounts for inclination in our correction; the PSF is always circular and so the effects of seeing are dependent on the projected inclination of the galaxy (for more details, see appendix C in G18). The scatter is also reduced for \vsigma{} on the right, but to a lesser extent. 

We note here that, when plotted as tracks, some of the tracks will move from being FR to SR as the seeing grows worse causing some to appear incomplete. It is important to divide these tracks in this way, as an observation of a regularly-rotating FR made in poor seeing may be observed as a SR \citep{Graham2019SDSS-IVGroups/clusters}. We need to ensure that these systems can also be corrected. 

In the lower panels of Figure \ref{fig:hf_correction}, we consider the SR observations of the N-body models. We show 383 individual tracks for both \lR{} and \vsigma. Of these, 49 observations remain within the SR regime across the full track length. Many more tracks begin at \psf{} > 0 in this plot, where the increased level of seeing has caused an intrinsically FR observation to drop below the criteria and appear as a SR. Nonetheless, we still bring all values back towards true while also reducing the scatter. A possible reason for this correction being effective for both FRs and SRs is that we have considered the effects of seeing in a relative parameter space. In this space the track shape is similar for both slow and fast rotation, as long as that rotation is regular. 

\begin{figure}
	\includegraphics[width=0.99\columnwidth]{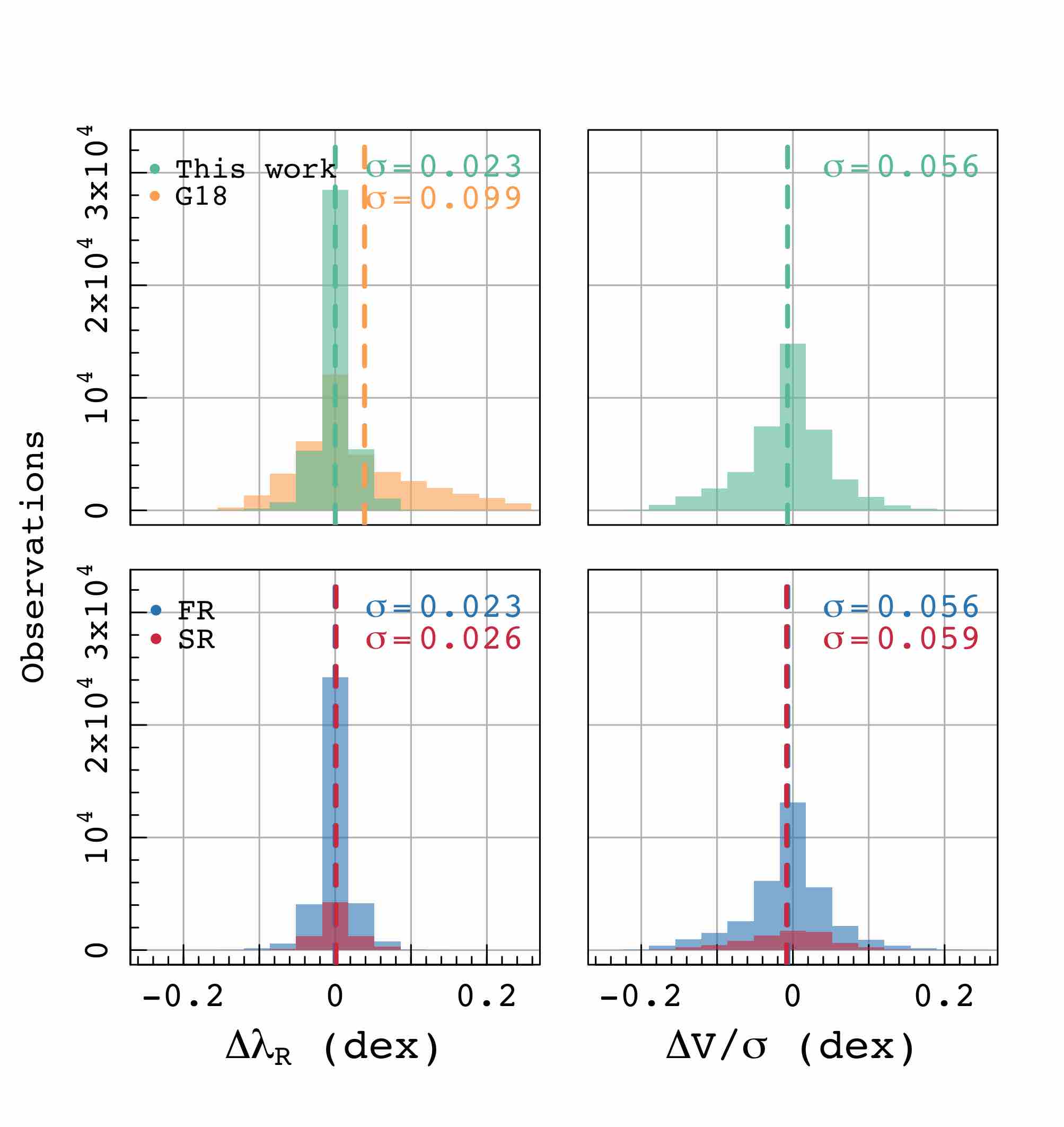}
    \caption{Histograms demonstrating the distributions of the corrected kinematics, \dlr{} (left) and \dvsig{} (right), for all 41,202/41,325 observations. (Above)  The green bars show \corrdlr{} and \corrdvsig{}; the orange bar in the right panel show the residuals following G18's correction. (Below) Breaking down the same sample into FRs and SRs in blue and red respectively. Dashed lines indicate the median and the $\sigma$ values describe the standard deviation of each distribution.}
    \label{fig:res}
\end{figure}

For the plots considering \lR{}, we compare the correction presented in this work to G18. In some respects, this comparison is unfair as the G18 correction was designed for FR measurements made within a single effective radius and for \sersic{} indices between $0.5 < n < 6.5$; hence, the distinction is made between those observations that meet the conditions and those that do not by colour coding \Reff{} $\neq$ 1 and $n > 6.5$ in yellow and valid G18 corrections in orange. This places emphasis on the fact that including a factor that fully parameterises galaxy shape is important if we wish to reduce the scatter in the \dlr{}-\psf{} space. 

The distribution of corrected values for the full sample of 41,202/41,325 observations are shown in Figure \ref{fig:res}. The following statistics are presented as the median of each distribution with the \nth{16} and \nth{84} percentiles below and above respectively ($\nu_{16\tiny{\text{th}}}^{84\tiny{\text{th}}}$). On average, the effect of seeing conditions is to reduce the values of \obsdlr{} $\sim -0.245_{\tiny{-0.388}}^{\tiny{-0.051}}$ and \obsdvsig{} $\sim -0.310_{\tiny{-0.463}}^{\tiny{-0.071}}$. By applying the correction presented in this work, these values become \corrdlr{} $\sim 0.000_{\tiny{-0.016}}^{\tiny{+0.017}}$ and \corrdvsig{} $\sim 0.000_{\tiny{-0.056}}^{\tiny{+0.038}}$. The key result is that we bring the median of the distribution back to zero, within \dlr{} and \dvsig{} $\sim 1 \times 10^{-3}$ dex. This is well below the statistical median uncertainty of \lR{} and \vsigma{} in surveys \citep{vandeSande2017TheSurveys}. We also significantly reduce the spread of the distribution in applying this correction. However, in comparing $\sigma$ for \corrdlr ($\sigma = 0.02$ dex) and \corrdvsig ($\sigma = 0.06$ dex), as shown in Figure \ref{fig:res}, we see that \lR{} is more effectively corrected than \vsigma{}. We believe that this is due to the fact that the seeing conditions impact the value of LOS velocity more than the dispersion; as \lR{} has factors of velocity in both the numerator and the denominator, these effects are partially cancelled out, unlike in \vsigma. The skew of each distribution demonstrates that we tend to under-correct our values. In comparison, the G18 correction has a more significant skew towards over-correction, where $\Delta \lambda_{R}^{\text{G18}}$ $\sim 0.006_{\tiny{-0.040}}^{\tiny{+0.129}}$.

In the lower panel of Figure \ref{fig:res}, we break down the distributions into samples of SRs and FRs in red and blue respectively. If we consider the effect of seeing on the rotator types independently, we find that \obsdlr{} $\sim -0.230_{\tiny{-0.374}}^{\tiny{-0.038}}$ for FRs and \obsdlr{} $\sim -0.330_{\tiny{-0.433}}^{\tiny{-0.209}}$ for SRs. Similarly, \obsdvsig{} $\sim -0.292_{\tiny{-0.462}}^{\tiny{-0.054}}$ for FRs and \obsdvsig{} $\sim -0.362_{\tiny{-0.463}}^{\tiny{-0.228}}$ for SRs. For both kinematic measurements, we see that FRs are relatively affected by seeing a lesser amount than SRs, as concluded in \cite{Harborne2019Alambda_R} because the median values are much larger in the SRs. However, following correction, we bring all values back towards zero successfully. On average, these corrected values have a distribution described by \corrdlr{} $\sim 0.000_{\tiny{-0.016}}^{\tiny{+0.015}}$ and \corrdvsig{} $\sim 0.000_{\tiny{-0.053}}^{\tiny{+0.036}}$ for the FRs; for the SRs, these values are \corrdlr{} $\sim -0.001_{\tiny{-0.020}}^{\tiny{+0.027}}$ and \corrdvsig{} $\sim 0.000_{\tiny{-0.065}}^{\tiny{+0.045}}$. The difference between the spread for the corrected FR and SR is very small, but the skews are opposite, with SRs more over-corrected on average than the FRs. However, while the effect of seeing on the SRs is relatively larger, the corrected position in real space is close to true due to the fact that these values are by definition smaller. 

\begin{figure*}
    \centering
    \includegraphics[width=\textwidth]{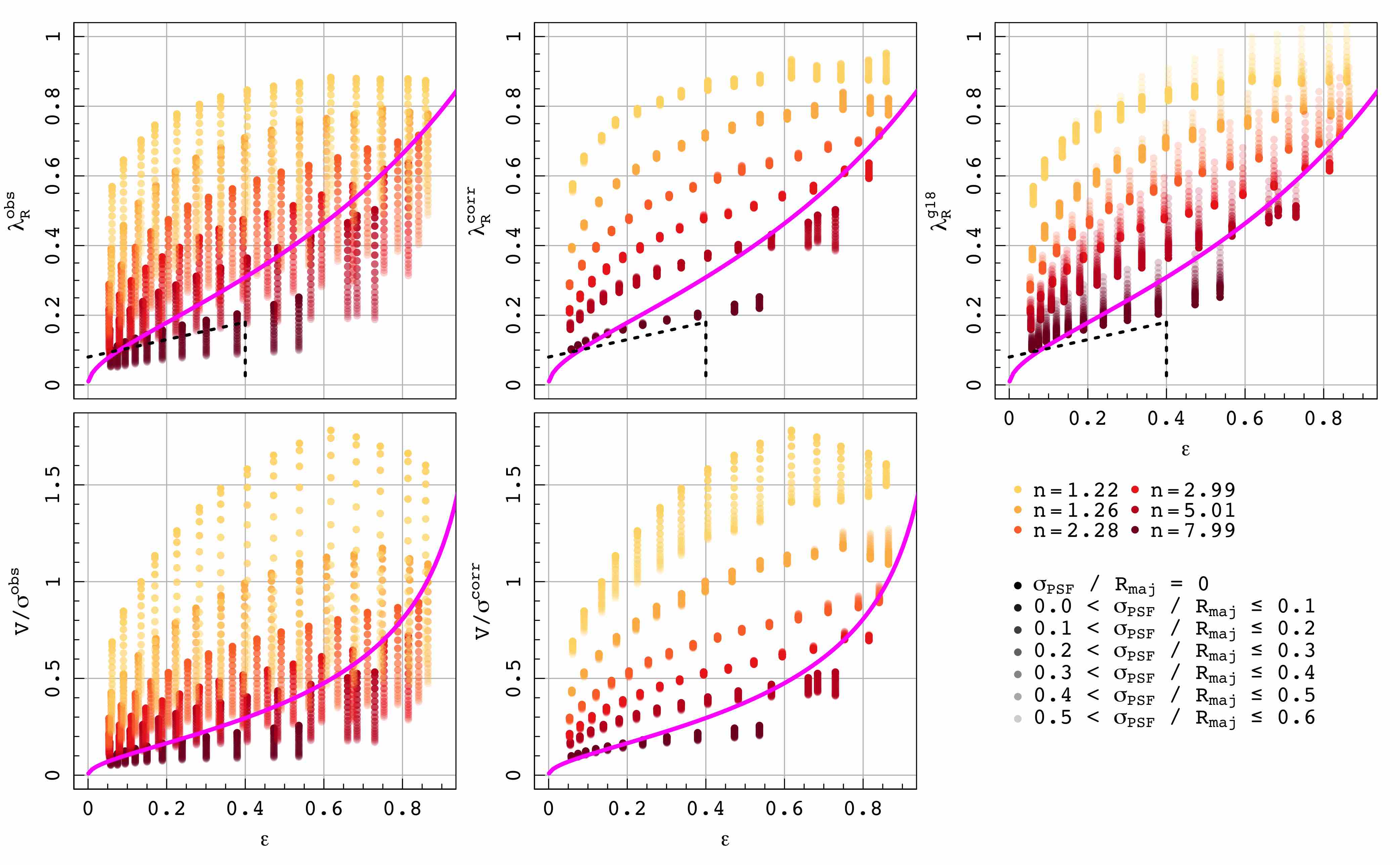}
    \caption{Showing the kinematic measurements for six galaxies made within 1 \Reff{} in the spin-ellipticity plane at a variety of seeing conditions up to \psf{} = 0.6. \lR{} (top) and \vsigma{} (bottom) are shown with respect to ellipticity, $\varepsilon$. The first column demonstrates the observed data. The centre column shows the same data following the application of the corrections in Equations \ref{eq:lr_corr} and \ref{eq:vsig_corr}. The final plot in the last column shows the data with the G18 correction applied. Colour denotes the \sersic{} index of the galaxy and opacity indicates the level of blurring applied. The magneta lines in all panels show the theoretical relation between spin and ellipticity for edge on galaxies, assuming $\delta \sim 0.7 \varepsilon$ (see \citeauthor{Cappellari2007TheKinematics} (\citeyear{Cappellari2007TheKinematics}) for more details). The dotted black lines in the top row show the FR/SR criteria from \citeauthor{Cappellari2016StructureSpectroscopy} (\citeyear{Cappellari2016StructureSpectroscopy}).}
    \label{fig:spin-ellipticity}
\end{figure*}

Kinematic measures are often presented as a function of ellipticity, where SRs and FRs can be distinguished. Figure \ref{fig:spin-ellipticity} demonstrates the effect of applying our correction to six galaxies from our sample in this spin-ellipticity plane for measurements made within 1 \Reff. On the left we demonstrate the observations made at a variety of seeing conditions up to the limit of \psf{} = 0.6, similar to the cuts made for observations in SAMI \citep{vandeSande2017TheSurveys}. As the conditions get poorer, we see that the measurements are shifted down towards the slow rotator regime. Following correction, these tails are significantly reduced for both \lR{} and \vsigma. 

Using the corrected panels in the centre of Figure \ref{fig:spin-ellipticity}, we can see a few of the deficiencies of equations \ref{eq:lr_corr} and \ref{eq:vsig_corr}. For \lR$^{\text{corr}}$, we see residual seeing effects present in the highly inclined systems. A similar effect is seen in the corrected \vsigma$^{\text{corr}}$, though this is secondary to the fact that the lowest \sersic{} index galaxy shown, n $=1.22$, has a large residual in comparison to the others. This is important to bear in mind when applying the \vsigma{} correction. We have not fully described all of the scatter within the \psf{} vs. \vsigma{} parameter space, leaving a residual that can be seen in the higher \vsigma{} values. This residual \sersic{} dependency is far less of an issue for \lR.

\begin{figure}
	\includegraphics[width=0.99\columnwidth]{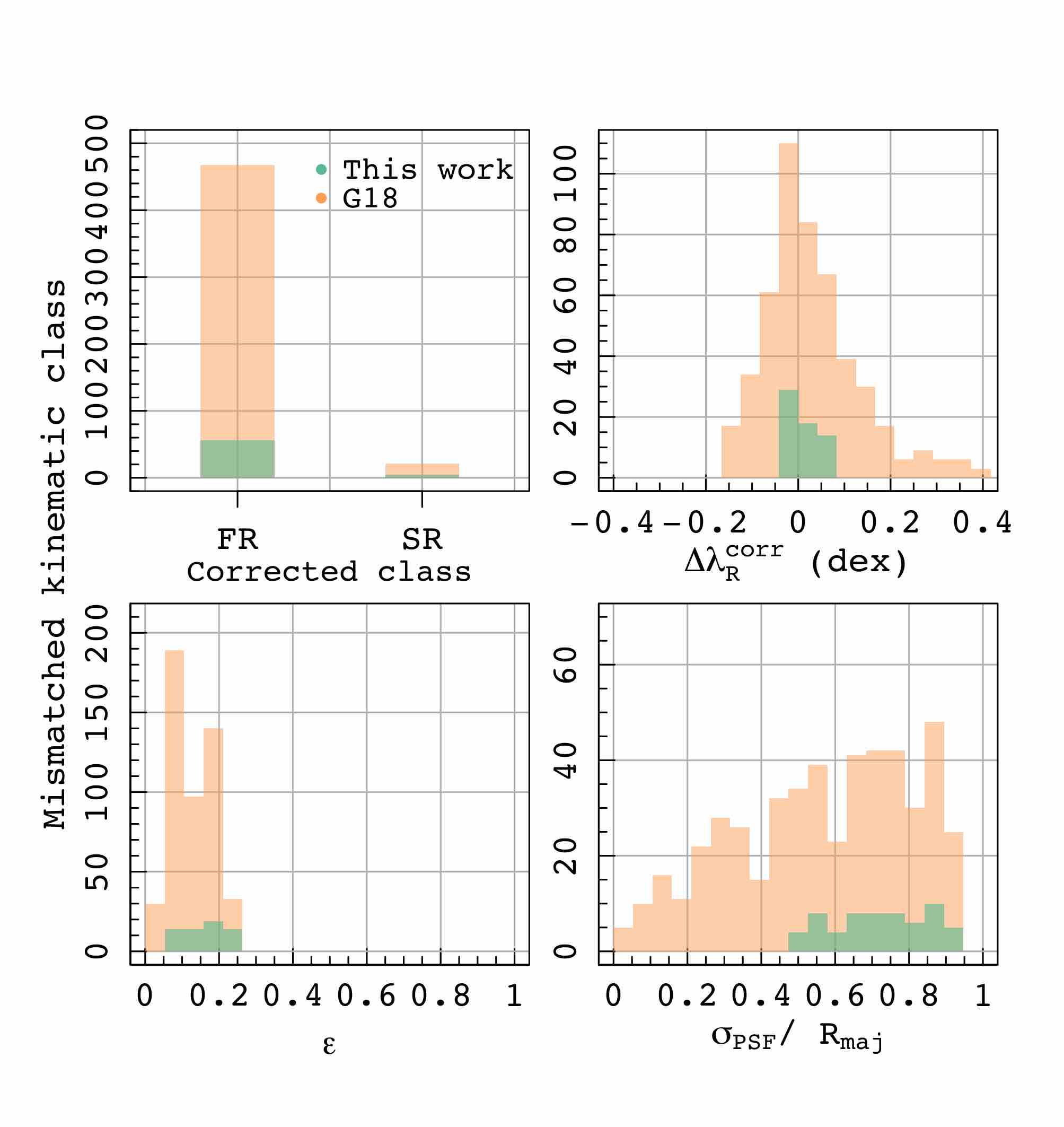}
    \caption{Histograms demonstrating the observations with mismatched kinematic classes following correction (61/489) by this work and G18 respectively, where we consider the true kinematic class to be that one defined based on the true \lR{} values measured within 1 \Reff, as explained in Section \ref{sec:FR/SR_class}.}
    \label{fig:mismatch}
\end{figure}

If we consider only valid observations (i.e. \lR{} and $\varepsilon > 0.05$) classified by their true kinematics, we found that 2.8\% of the full sample of 41,202 observations are intrinsic SR (as shown in Figure \ref{fig:observations}). Redoing our classification of FRs and SRs using the measured and corrected \lR{}, 2.7\% fall into the SR regime (corrected from a fraction of 17\%). This brings us much closer to the 2.8\% of the original, perfect-seeing classifications. We correct  99.9\% (all but 61 observations) back to their true classification. These remaining mismatched observations are nearly all SR that have been observed to have spin parameters close to the \lR{} = 0.05 cut-off. These observations tend to occupy the 0.5 < \psf{} < 1 range, and have observed ellipticities, $\varepsilon < 0.2$. The residuals between the true parameter and the corrected values are less than 0.05, but even this level of difference causes the incorrect galaxy classification to be assigned. Overall, however, this difference is very small. We show the distributions of mismatched kinematic class observations in Figure \ref{fig:mismatch}. By comparison, the G18 correction reduces this fraction to 1.7\%.

\subsection{EAGLE galaxy results}
\label{sec:res_eagle}
The correction presented in this work has been derived using a set of N-body regularly-rotating models. It is important to verify that this correction is valid also for an independent set of galaxies with different assembly histories. Furthermore, while the majority of galaxies appear as FRs, as discussed in Section  \ref{sec:method}, it is important to understand the effect this correction has on the full data set i.e. including the irregular rotators. We have selected three FR and four SR galaxies that exhibit slow or irregular rotation from the \eagle{} simulation for this test, as outlined in Table \ref{tab:eagle}, and present their analysis below. 

\begin{figure*}
	\includegraphics[width=\textwidth]{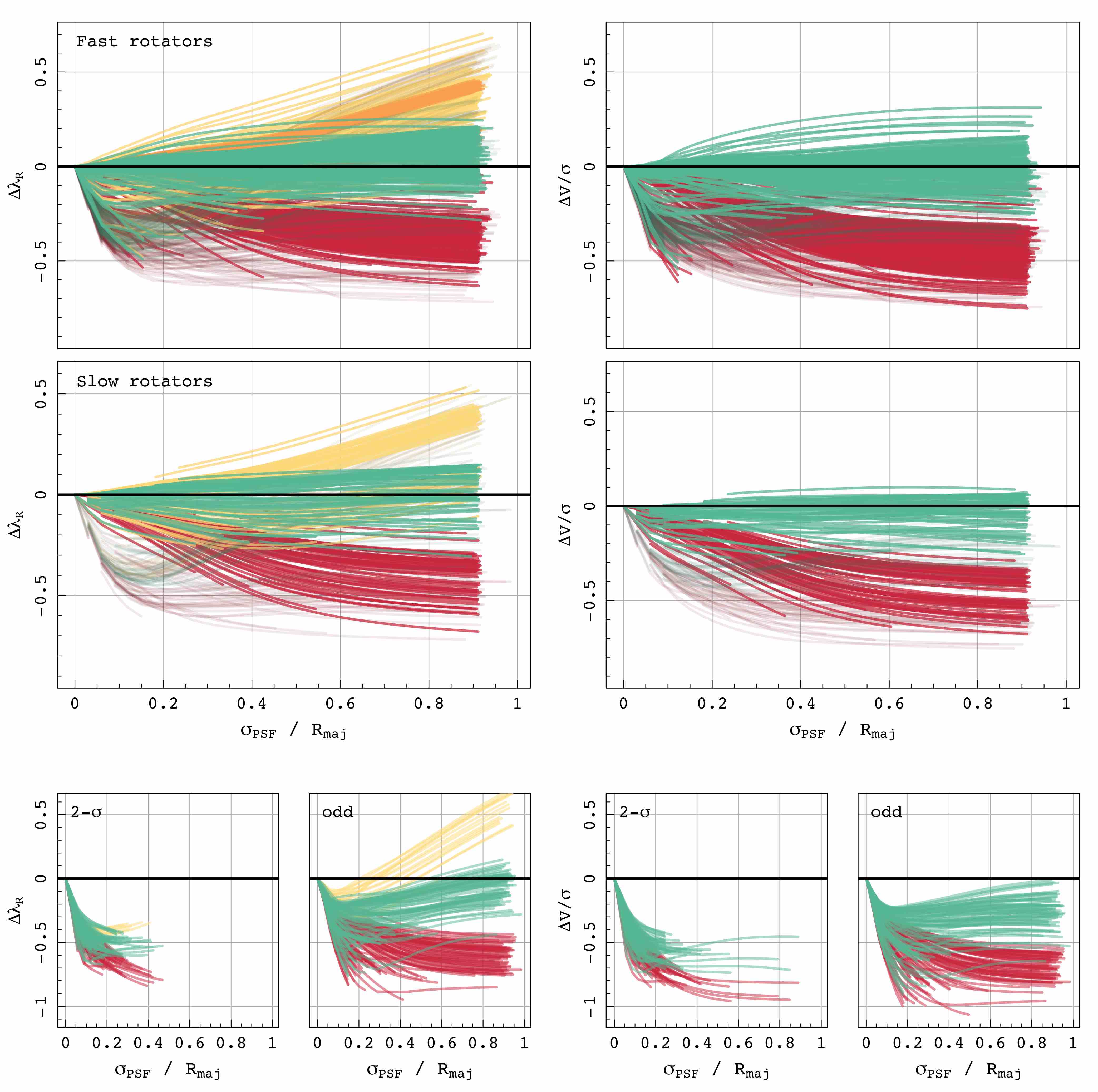}
    \caption{Considering the effect of seeing on the kinematics of the \eagle{} galaxies demonstrating the FR sample (top), the SR sample (centre) and the two irregular examples (bottom). The lines are coloured as in Figure \ref{fig:hf_correction}. We have shown tracks of systems with higher particle resolution in the opaque, thicker lines. Transparent, darker lines show tracks for the low particle resolution systems. The 2-$\sigma$ and odd galaxies shown below also have low particle resolution. The success of the correction seems to show strong dependence on resolution.}
    \label{fig:EAGLE_SR}
\end{figure*}

We begin with 20,615 observations of the \eagle{} galaxies listed in Table \ref{tab:eagle}. When we remove any observations using the cuts explained in Section \ref{sec:method-correct} ($\varepsilon$ and \lR{} or \vsigma{} < 0.05), this reduces the sample size to 15090 observations. Using the criteria in Equation \ref{eq:FRcriteria}, we divide the sample of regularly rotating \eagle{} observations into 9722/10100 FRs and 2440/2497 SRs for \lR{} and \vsigma{} respectively. We also have two galaxies that have been identified as having irregular kinematic morphology (i.e. \texttt{GalaxyIDs = 10770392} (2-$\sigma$) and \texttt{10048611} (odd)) which are analysed separately.

Plotting these in Figure \ref{fig:EAGLE_SR}, we have 448 FR \psf{} tracks and 102 SR tracks, of which 8 observations are fully SR across all seeing conditions. At first glance, this set of galaxies shows a broader range of scatter in both the observed and corrected track shapes than the N-body set shown in Figure \ref{fig:hf_correction}. 

The largest discrepancy between the N-body and \eagle{} samples that may cause these effects is the particle resolution of each simulation. The three regular FRs have an order of magnitude fewer particles than the full N-body catalogue of galaxies (N$_{\text{EAGLE}}\sim 1\times10^5$ vs. N$_{\text{N-body}} = 2.5\times10^6$). The four galaxies chosen for their SR and irregular properties have an order of magnitude fewer particles again (N$\sim 1\times10^4$). From \cite{Ludlow2019EnergySizes}, we know that for galaxies with low particle numbers, two-body particle scattering will affect the size and very likely the velocity dispersion profile of a galaxy, causing the disks to have larger scale heights than expected. This will also be affected by the intrinsic softening size and the temperature floor within \eagle. While we have selected a conservative particle limit based on convergence tests in \cite{Lagos2017AngularEAGLE}, this value may need to be even larger for generating mock observations using tools such as \simspin. We know that the softening in the simulation alone (0.70 physical kpc) is equivalent to a level of blurring which contributes on top of the effects we have added, meaning that the value we have taken as $\Delta \lambda_R^{\text{true}}$ is more like observations of the $N$-body systems made at \psf{} > 0. This will lead to an increased level of scatter in our corrected kinematics as we are applying this formula without accounting for the added seeing contribution. 

The particle resolution is also very important to the final track shape that is observed, as highlighted in Figure \ref{fig:EAGLE_SR}. In the top and centre panels, we have highlighted the higher resolution models in a darker, thicker colour than the low resolution set. The irregular tracks shown in the lower panels are also of low resolution particle models. Noticeably, the tracks for the low resolution models follow a different shape to the higher resolution models, appearing to reduce rapidly at very small seeing effects and plateau to a maximum $\Delta$ earlier in the track. The convolution of any size PSF with the sparser particle numbers will make it much easier for atmospheric effects to blur out any rotation. 

This also manifests in the shape of the track for the smaller apertures (\Reff-factor = 0.5) which, by definition, contain less than half the total number of particles in the model. We see that these tracks show a greater departure from the expected ``S''-shaped track, even for the FR selected simulations with $1\times10^5$ particles. 

Because these low resolution models have the highest particle numbers available from the SR sample extracted from the \eagle{} \texttt{RefL0100N1504} box, we cannot fully evaluate how effective this correction is for irregular systems. Nevertheless, we show the 35 \psf{} tracks for 395/503 observations of \lR{} and \vsigma{} respectively in the lower panels of Figure \ref{fig:EAGLE_SR} for completeness. Interestingly, we see that using the criteria in Equation \ref{eq:FRcriteria}, there are several of the observations of the irregular galaxies that would have been classified as FRs at low seeing (as can be seen by tracks beginning between $0 <$ \psf{} $< 1$ in Figure \ref{fig:EAGLE_SR}). This may suggest that looking for irregular systems within the SR category may not be sufficient. In the future, we would like to test this on our own cosmological zoom models of irregular systems built at a much higher particle resolution, but this is beyond the scope of this paper. 

\begin{figure}
	\includegraphics[width=0.99\columnwidth]{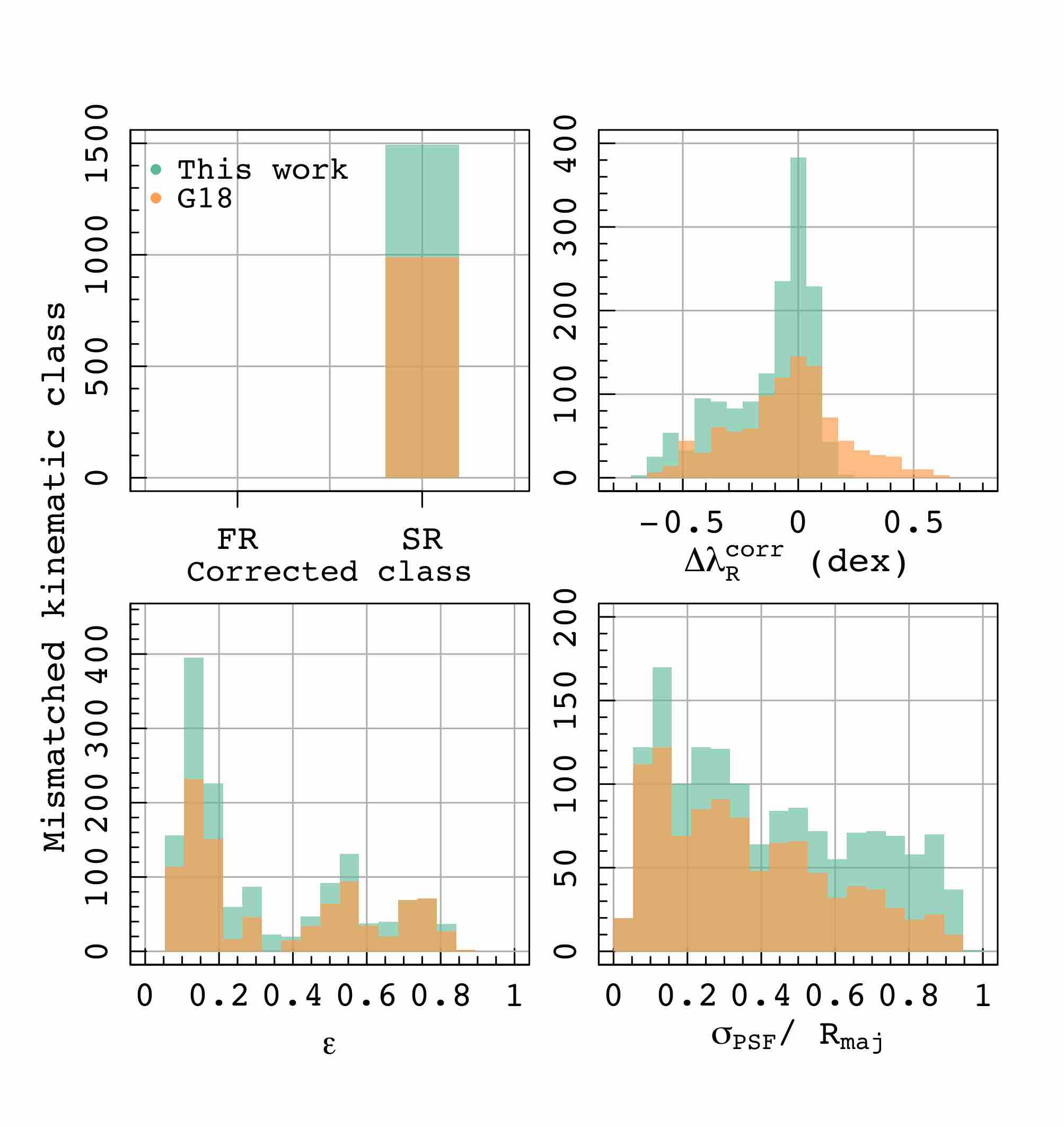}
    \caption{Histograms demonstrating the observations with mismatched kinematic classes following correction (1494/990) by this work and G18 respectively for the \eagle{} galaxies, where we consider the true kinematic class to be that one defined based on the true \lR{} values measured within 1 \Reff, as explained in Section \ref{sec:FR/SR_class}.}
    \label{fig:mismatch_eagle}
\end{figure}

\begin{figure*}
	\includegraphics[width=\textwidth]{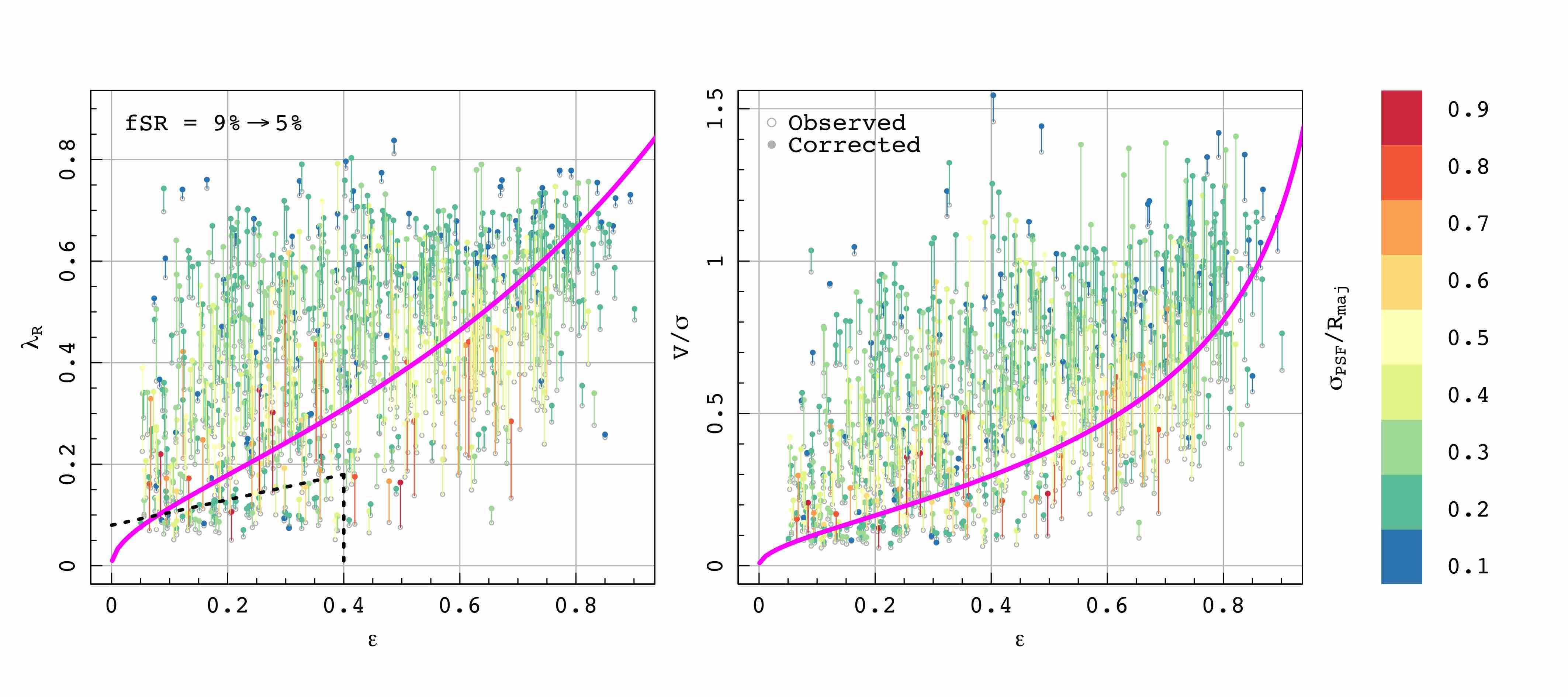}
    \caption{The effect of applying our corrections to the SAMI DR2 data. (Above) The \lR{}-$\varepsilon$ plane, with magenta and dash lines as in Figure \ref{fig:spin-ellipticity}, and (below) the \vsigma{}-$\varepsilon$ plane for the range of data that meets the quality criteria cuts described in the text. Grey empty points demonstrate observed kinematics and full coloured points show their corrected kinematic values. Lines connect each point to demonstrate how far each point has moved. The colours show the size of the PSF ($\sigma$) as compared to the effective radius of the galaxy. In the upper left hand corner, we show how the SR fraction has changed in the sample before and after correction.}
    \label{fig:SAMI_DR2}
\end{figure*}

We consider the average statistics of the distributions using the median and \nth{16} and \nth{84} percentiles below and above respectively. For the FR observations, \obsdlr{} $\sim -0.275_{\tiny{-0.413}}^{\tiny{-0.055}}$ reduces to \corrdlr{} $\sim -0.013_{\tiny{-0.127}}^{\tiny{+0.091}}$; SR observations, \obsdlr{} $\sim -0.370_{\tiny{-0.517}}^{\tiny{-0.181}}$ reduces to \corrdlr{} $\sim -0.027_{\tiny{-0.239}}^{\tiny{+0.079}}$; and for the irregular galaxies, \obsdlr{} $\sim -0.500_{\tiny{-0.651}}^{\tiny{-0.263}}$  reduces to \corrdlr{} $\sim -0.208_{\tiny{-0.442}}^{\tiny{-0.055}}$. The overall effect of the correction does move the kinematic measurement up and closer to true, but does not do much to reduce the spread. Values are also often skewed towards under correction; this is understandable if we consider that the effect of particle softening in \eagle{} effectively moves what we have taken as the intrinsic value along the track towards \psf{} > 0. However, even in the worst case of the irregular systems, we reduce the effects of seeing by a factor of 2. For the majority of the regularly rotating systems drawn from \eagle{}, we are reducing the effects by a factor of 4. 

By applying our correction to this data set, the SR fraction is reduced from 19\% to 11\% (where the fraction for the intrinsic data is 1\%). In this case, the G18 correction reduces the SR fraction to 7\%. In Figure \ref{fig:mismatch_eagle}, we consider the observations that have been labelled incorrectly following correction. We find that the distributions for this work and G18 are fairly similar to one another. From Figure \ref{fig:res}, we know that the G18 correction tends to over-correct systems pushing them further toward the FR regime. Given that all mismatched observations have been labeled as SRs after correction implies that neither our correction nor G18 is pushing these observations far enough, but that G18 goes further, as can also be seen in the residual histogram in the upper right panel of Figure \ref{fig:mismatch_eagle}. While we incorrectly categorise more observations as SR, we are getting the corrected \lR{} values much closer to true than G18, which has a tail of observations that are strongly over-corrected.  

Overall, this independent test confirms that this correction is applicable to a broad number of regularly-rotating galaxies. The issue of particle resolution makes it difficult to conclude how effective this correction is for irregularly-rotating systems. However, for the systems shown in Figure \ref{fig:EAGLE_SR}, we see that the net effect is to move them closer to true. 

It seems reasonable to assume that the \psf{} track shape with seeing is different for a galaxy with embedded, irregular kinematic features. However, if the impact of seeing is large enough to have washed out these features, we expect to move the kinematics on a similar track, given that the tracks for FRs and SRs are similar. In applying this correction to irregulars, at least in the cases shown in Figure \ref{fig:EAGLE_SR}, the result is moving the value towards the correct value. 

\begin{figure*}
	\includegraphics[width=0.8\textwidth]{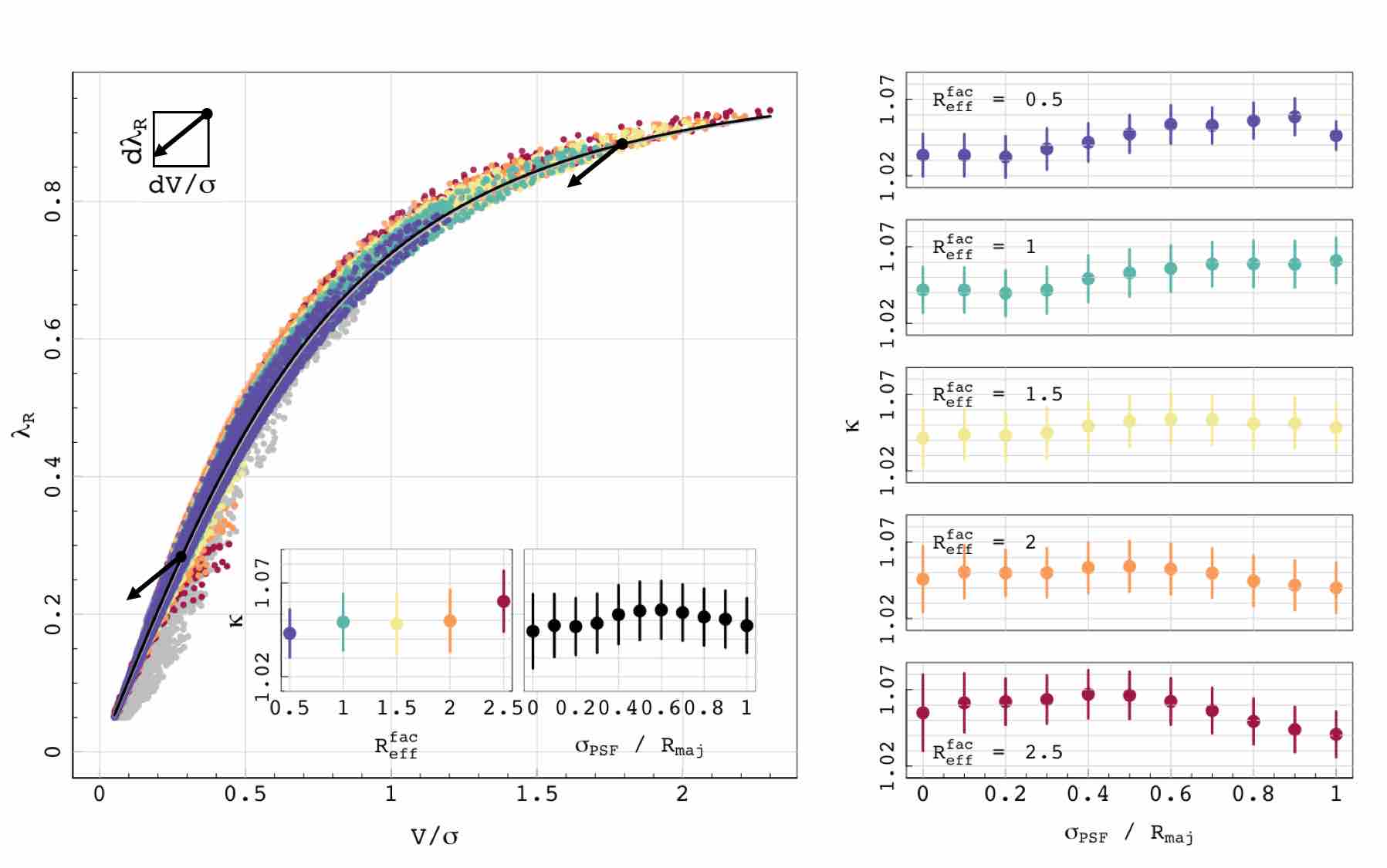}
    \caption{Demonstrating the relationship between \lR{} and \vsigma. We have fitted Equation \ref{eq:kappa} to our observations. The coloured points show the N-body observations, with colour highlighting the measurement radius; grey points represent all \eagle{} observations. The two inbuilt panels on the left demonstrate the $\kappa$ values fitted to the whole sample binned by \Reff{} or PSF (left and right respectively). The trends of $\kappa$ with PSF within each measurement radius group are shown on the right. Errors shown are the standard deviation of the residual sum of least square deviations. The arrow annotations in the upper left corner and along the distribution explain the trends shown on the right. All three arrows have the same gradient.}
    \label{fig:kappa}
\end{figure*}

\subsection{Application to real data}
\label{sec:res_DR2}

We have demonstrated that these corrections are applicable to a whole range of different simulated data sets, thus it seems prudent to explore the effect of this correction on real observations. Here, we apply our corrections to SAMI DR2 observations \citep{Scott2018TheProducts}. Following the methodology of \cite{vandeSande2017TheKinematics, vandeSande2019TheSimulations}, we begin by applying the standard quality cuts to the sample. From the initial sample of 960 galaxies, we exclude 11 observations for which the radius out to which the stellar kinematics can be accurately measured is less than the half-width half-maximum of the PSF (HWHM$_{\text{PSF}}$), and a further 95 observations which are poorly resolved for which kinematics can not be obtained. Galaxies are removed that have a stellar mass less than $10^{9.5}$ M$_{\odot}$ due to the reliability of kinematics below this mass, removing 89 systems from the sample. We further remove 24 galaxies with $\varepsilon$, \lR{} or \vsigma{} < 0.05 and \psf{} > 1, in line with our empirical correction requirements. This leaves us with 741 observations from DR2. In order to demonstrate the effect of the correction, we compare the SR fraction of this sample before and after correction. We find, using the FR/SR criteria of \cite{Cappellari2016StructureSpectroscopy} in Equation \ref{eq:FRcriteria} that this sample of 914 galaxies contains $9\%$ SR. Using Equations \ref{eq:lr_corr} and \ref{eq:vsig_corr}, we calculate the corrected kinematic values, \lR{} and \vsigma, and re-calculating the SR fraction, we find that this has dropped to $5\%$.

In Figure \ref{fig:SAMI_DR2}, we show these observations before and after correction in the spin-ellipticity plane. The grey empty points show the original observed kinematic parameter, while the full, coloured points show the value after correction. Each pair of points is joined by a line to show how far that single observation has moved following correction. The sizes of these lines vary across the parameter space, which we have shown is dependent on a combination of galaxy shape, ellipticity, measurement radius and seeing conditions.

The range of colours in this figure demonstrates that such corrections are important for comparing survey data. While the predominant \psf{} value is $\sim0.3$, this value varies from $0.08 - 0.82$. The length of lines connecting uncorrected and corrected points does not scale linearly with \psf, so simple adjustments made using the seeing conditions alone are not sufficient to make consistent comparisons. This justifies the requirement in the community for the corrections presented in this work. We further find that the corrections presented in this paper are important for quantifying the presence of a bimodality in the \lR-$\varepsilon$ plane (van de Sande et al., in prep).
 
\subsection{The relationship between kinematic parameters}
\label{sec:kappa}
Finally, we use our idealised data set, as shown in Table \ref{tab:sims}, to examine how seeing conditions and measurement radius affect the relationship between \lR{} and \vsigma. In \cite{Emsellem2007TheGalaxies}, they introduce a simple approximation that allows conversion from one kinematic parameter to the other:
\begin{equation}
    \lambda_R \approx \frac{\kappa \left(V/\sigma\right)}{\sqrt{1 + \kappa^2\left(V/\sigma\right)^2}}.
    \label{eq:kappa}
\end{equation}
This $\kappa$-parameter has been calculated and compared across different surveys (\textsc{Sauron} $\sim 1.1$ \citep{Emsellem2007TheGalaxies},  \textsc{Atlas}$^{\text{3D}} \sim 1.06$ (\citeauthor{vandeSande2017TheSurveys} \citeyear{vandeSande2017TheSurveys}, \citeauthor{Emsellem2011TheRotators} \citeyear{Emsellem2011TheRotators} found a value of $\kappa \sim 1.1$ originally with mixed aperture sizes), \textsc{SAMI} $\sim 0.97$ \citep{vandeSande2017TheSurveys}), but the value and the level of scatter in the plotted distribution is not consistent. The \textsc{SAMI} distribution of \lR{} vs. \vsigma{} is a much tighter distribution than that seen in \textsc{Atlas}$^{\text{3D}}$, for example.\footnote{The \textsc{SAMI} value is also smaller due to the different definition of \lR{} that is used in this survey. For more details on this difference, see Appendix \ref{sec:app_lre}.} 

In \cite{vandeSande2017TheSurveys}, they demonstrate that the scatter in the distribution is affected by systematics such as the difference in spatial resolution and seeing conditions. \textsc{Atlas}$^{\text{3D}}$ has a much higher resolution, such that complex internal dynamical features are not washed out by atmospheric blurring. It follows that there is a larger scatter in the $\kappa$-relation for this survey, as \lR{} is designed to better distinguish internal kinematic structures than \vsigma{} \citep{Emsellem2007TheGalaxies, Emsellem2011TheRotators}. 

In \cite{Cortese2019ThePhase}, it was suggested that conversion between \lR{} and \vsigma{} may not be trivial if $\kappa$ has a strong dependence on the size of the PSF. We can use our data set to investigate this. We measure $\kappa$ in our idealised data set using regression, minimising the sum of least squares for the uncorrected values of \lR{} and \vsigma. We use the uncorrected values because we want to examine how $\kappa$ is affected by seeing. The error on this fitted value can be found by, 
\begin{equation}
    \hat{\sigma} = \sqrt{\frac{Q}{n - p}},
\end{equation}
where $Q$ is the remaining sum of the square residuals from the best fit $\kappa$ value, $n$ is the number of observations in the data set, and $p$ is the number of parameters being fit (i.e. $p=1$ for the one parameter, $\kappa$). In the case of the full 41,231 observations, $\kappa = 1.05 \pm 0.016$. This is shown in Figure \ref{fig:kappa} by the black curve.

Our idealised observations have been generated using the same resolution as the \textsc{SAMI} survey and so we cannot investigate the effect of spatial resolution with this data alone. Furthermore, we have only considered regular rotators in this analysis. When we add our sample of observations from \eagle{} to the distribution, we find they do add to the overall scatter of the distribution as shown by the grey points in Figure \ref{fig:kappa}. When fitting $\kappa$ to the full idealised plus \eagle{} data set, we find that the value decreases slightly, but with larger scatter, to $\kappa = 1.04 \pm 0.018$. This is shown by the grey curve in Figure \ref{fig:kappa}. This indicates that $\kappa$ value may also see variations due to the number of irregular systems in the data-set. 

We then bin our idealised data set of 41,202 observations by the five \Reff{}-factors and by seeing conditions, with bin widths of $\Delta$ \psf{} = 0.1. For each bin, we fit Equation \ref{eq:kappa}. The inbuilt panels on the left demonstrate the trends of $\kappa$ with these variables independently.

As in \cite{vandeSande2017TheSurveys}, we find that there is a slight effect due to seeing, $\Delta \kappa \sim \pm 0.02$, though there is a stronger dependence on aperture size, where we see a positive correlation with measurement radius. Looking at the seeing conditions within each measurement radius bin, we see that the correlations with seeing conditions change with the measurement radius being considered. These trends are shown in Figure \ref{fig:kappa} on the right. 

The gradient change can be understood by considering the trends seen in Figure \ref{fig:psf_dlr_dvsig}. We know that, overall, the effect of seeing on \vsigma{} is slightly greater than that of \lR{}. We also know this effect is consistent across R$_{\text{eff}}^{\text{fac}}$, but on average we measure higher values of \lR{} and \vsigma{} at larger measurement radii. At the lower spin end of the parameter space, a more rapid change in \vsigma{} is going to scatter points towards higher $\kappa$. Moving points in a similar direction at higher spin, the effect causes $\kappa$ to be pulled down. This is illustrated by the annotations in Figure \ref{fig:kappa}, where each arrow has exactly the same gradient. The absolute change in $\kappa$ seen in Figure \ref{fig:kappa} is very small, with the maximum variation $\Delta \kappa \sim \pm 0.02$ in all figures on the right-hand side. This may be due to the fact that there is opposing trends with \sersic{} index and ellipticity for \lR{} and \vsigma{} that we see in Figure \ref{fig:sigmoid_fit_residuals}. The potential variation in the gradient change with \psf{} at each R$_{\text{eff}}^{\text{fac}}$ is important. If a given survey contains a larger portion of one \Reff-factor than another, the average fit to the full data set will be biased. This also confirms the results of \cite{vandeSande2017TheSurveys}, in so much as reiterating the need for such aperture corrections. 

\section{Discussion and Conclusions}
\label{sec:conclusions}
In this work, we have investigated how seeing conditions affect kinematic measurements of simulated galaxies across a range of morphologies. We find that atmospheric blurring causes the number of SRs inferred to be artificially increased relative to its intrinsic value, which can cause problems when trying to infer trends in the kinematic morphology-density relation. In our sample of regular rotators, the effects of seeing cause the fraction of SR to increase from 3\% to 17\%. 

This has led us to design a new empirical correction, which can be applied to reduce these effects. For our full sample of regular rotators, we see that the SR fraction for corrected values is returned to 3\%, and the corresponding spread of the distribution is \corrdlr{} $\sim -0.000_{\tiny{-0.024}}^{\tiny{+0.030}}$ dex and \corrdvsig{} $\sim -0.002_{\tiny{-0.037}}^{\tiny{+0.036}}$ dex. While our irregular sample has issues relating to the resolution of the models considered, we have demonstrated that this formula brings observed values of \corrdlr{} back to within $\sim -0.208_{\tiny{-0.442}}^{\tiny{-0.055}}$ dex for the irregular systems. We expect this offset to be due to the galaxy models rather than the ability of the correction and hence these values should improve with a better test sample, such as cosmological zoom simulations with higher particle resolution. We advise caution when using models with ``low'' ($N \sim 10^5$) particle numbers for kinematic measurements. 

\bigskip

Our formulae successfully corrects the \lR{} and \vsigma{} measurements of both fast and slow rotators, but still has the following limitations:
\begin{enumerate}
    \item We have designed this correction based on a range of fast regular-rotators with \sersic{} indices from $1 < n < 5.5$. We also require that \lR{}, \vsigma{} and ellipticity values are larger than 0.05. We have then shown that this correction is valid for a large variety of regular rotators with $n \leq 8$, whether those observations are classed SRs or FRs. Confirming the effectiveness of these formulae outside of this range, or for systems that show irregular kinematic morphology, is beyond the scope of this paper.
    \item We have made the assumption that the \sersic{} index, projected inclination and seeing PSF have been accurately measured and do not propagate errors in these parameters through our equation.
    \item We have also assumed that the scatter in the parameter space is fully described by shape, inclination, seeing conditions and measurement radius. 
\end{enumerate}

With respect to (i), this is an unfortunate limitation of all kinematic corrections. For systems in which the kinematically distinct features are washed out by seeing or resolution, we cannot hope to recover them using an empirical correction. Similarly, finite numerical resolution - an issue for the EAGLE sample of simulated galaxies - cannot be corrected for. Nevertheless, we have shown that such a correction is reasonably effective for a small sample of slow regularly-rotating systems and that the effect on the \eagle{} irregulars is to move them closer to their true value. Therefore, we believe Equations \ref{eq:lr_corr} and \ref{eq:vsig_corr} could be very useful for correcting all data in a statistical sense. Furthermore, we have shown that in the relative parameter space, the tracks for regular FRs and SRs are similar and so this correction works in both cases. We see the average effect of seeing reduced from \obsdlr{} $\sim -0.241_{\tiny{-0.382}}^{\tiny{-0.040}}$ to \corrdlr{} $\sim -0.001_{\tiny{-0.021}}^{\tiny{+0.029}}$ for FRs and \obsdlr{} $\sim -0.337_{\tiny{-0.454}}^{\tiny{-0.203}}$ to \corrdlr{} $\sim -0.002_{\tiny{-0.037}}^{\tiny{+0.036}}$ for SRs; 
equivalently, \obsdvsig{} $\sim -0.306_{\tiny{-0.472}}^{\tiny{-0.061}}$ to \corrdvsig{} $\sim -0.003_{\tiny{-0.077}}^{\tiny{+0.033}}$ for FRs and \obsdvsig{} $\sim -0.372_{\tiny{-0.497}}^{\tiny{-0.230}}$ to \corrdvsig{} $\sim -0.013_{\tiny{-0.110}}^{\tiny{+0.040}}$ for SRs. 

In making the assumption that other galaxy properties can be accurately calculated, we have ignored a significant factor of uncertainty that will result in scatter in the corrected kinematics. As discussed in \cite{Graham2018SDSS-IVProperties}, effective radii are difficult to measure in a consistent and accurate manner and the error in $\sigma_{\text{PSF}}$, depending on the survey, is $\sim10\%$ \citep{Law2016THESURVEY, Allen2015TheRelease}. We have not accounted for signal-to-noise in this work, and suggest that this addition may add to the uncertainty. Furthermore, \sersic{} indices have a minimum uncertainty that can be inflated by resolution, improper modelling of the sky, or PSF, and we found assigning a single component value to two component models was difficult. The combination of all of these issues makes the error in the corrected kinematics impossible to quantify accurately. Hence, we encourage conservative estimates of the corrected values.

Finally, we believe that the latter assumption (iii) - that scatter within the parameter space can be fully described using the factors we have considered - been shown to be valid throughout Section  \ref{sec:method-correct}, especially for \lR. However, we have also established that the magnitude of \lR{} is decreased much less by seeing conditions than \vsigma{} and that it is much easier to account for these effects through the use of our correction (where the standard deviation for $\Delta \lambda_R^{\text{corr}} \sim0.02$ versus $\Delta \text{V}/\sigma^{\text{corr}} \sim0.06$). We have noticed a shift in the community towards using \vsigma{} over \lR{} in recent years \citep{vandeSande2019TheSimulations, Cortese2019ThePhase}, but put forward the suggestion that \lR{} is more robust and comparable across different surveys once corrected. We also note here that we have not considered the spatial sampling within these equations, but expand on this further in Appendix \ref{app:ch5_spaxels}. The effect of spatial sampling is complex and multi-dimensional, but we demonstrate that it is important to consider when correcting the kinematics of small systems observed with poor seeing.

Future surveys will gain large number statistics of kinematic observations. We present these formulae as a way to correct the distribution of \lR{} and \vsigma{} in a manner that is not biased towards specific shapes or seeing conditions. Making such comparisons on an even footing is vitally important to advances within the field of galaxy evolution. 

\section*{Acknowledgements}

KH is supported by the SIRF and UPA awarded by the University of Western Australia Scholarships Committee. JvdS acknowledges funding from Bland-Hawthorn's ARC Laureate Fellowship (FL140100278) and support of an Australian Research Council Discovery Early Career Research Award (project number DE200100461) funded by the Australian Government. LC is the recipient of an Australian Research Council Future Fellowship (FT180100066) funded by the Australian Government. CP acknowledges the support of an Australian Research Council (ARC) Future Fellowship FT130100041, the support of ARC Discovery Project DP140100198 and ARC DP130100117. 

This research was conducted under the Australian Research Council Centre of Excellence for All Sky Astrophysics in 3 Dimensions (ASTRO 3D), through project number CE170100013. CL is also directly funded by ASTRO 3D. This research was undertaken on Magnus at the Pawsey Supercomputing Centre in Perth, Australia. 

This research uses public data from the SAMI Data Release 2 and we thank the team for their work on this survey. The Sydney-AAO Multi-object Integral field spectrograph (SAMI) was developed jointly by the University of Sydney and the Australian Astronomical Observatory. The SAMI Galaxy Survey is supported by the Australian Research Council Centre of Excellence for All Sky Astrophysics in 3 Dimensions (ASTRO 3D), through project number CE170100013, the Australian Research Council Centre of Excellence for All-sky Astrophysics (CAASTRO), through project number CE110001020, and other participating institutions. The SAMI Galaxy Survey website is \url{http://sami-survey.org/}.

We also acknowledge the Virgo Consortium for making their simulation data available. The \eagle{} simulations were performed using the DiRAC-2 facility at Durham, managed by the ICC, and the PRACE facility Curie based in France at TGCC, CEA, Bruy\`{e}resle-Ch\^{a}tel. 

\section*{Data Availability Statement}
The SAMI data used in this work is publicly available and can be accessed online via the Australian Astronomical Optics Data Central. See the SAMI website for access details and instructions: \url{https://sami-survey.org/news/sami-data-release-2}. Similarly, the \eagle{} data used in this work is publicly available at \url{http://virgodb.dur.ac.uk/}. The other data products underlying this article, such as the N-body galaxies used, will be shared on reasonable request to the corresponding author. 

%%%%%%%%%%%%%%%%%%%%%%%%%%%%%%%%%%%%%%%%%%%%%%%%%%



\bibliographystyle{mnras}
\bibliography{references} % if your bibtex file is called example.bib

%%%%%%%%%%%%%%%%%%%%%%%%%%%%%%%%%%%%%%%%%%%%%%%%%%

%%%%%%%%%%%%%%%%% APPENDICES %%%%%%%%%%%%%%%%%%%%%

\appendix
\section{For \texorpdfstring{$\lambda_R$}{Lr} measured using elliptical radii}
\label{sec:app_lre}

When measuring the observable spin parameter, $\lambda_R$, the radial parameter associated with a spaxel, $R_i$, in Equation \ref{eq:lambdaR} can be considered as the radius of a circle that passes through the relevant spaxel. Alternatively, the radius can be defined as the semi-major axis of an ellipse that would pass through the spaxel. This is the common method used within the \textsc{Sami} team \citep[e.g.][]{Cortese2016TheMorphology, vandeSande2017TheKinematics}. 

\begin{figure}
	\includegraphics[width=\columnwidth]{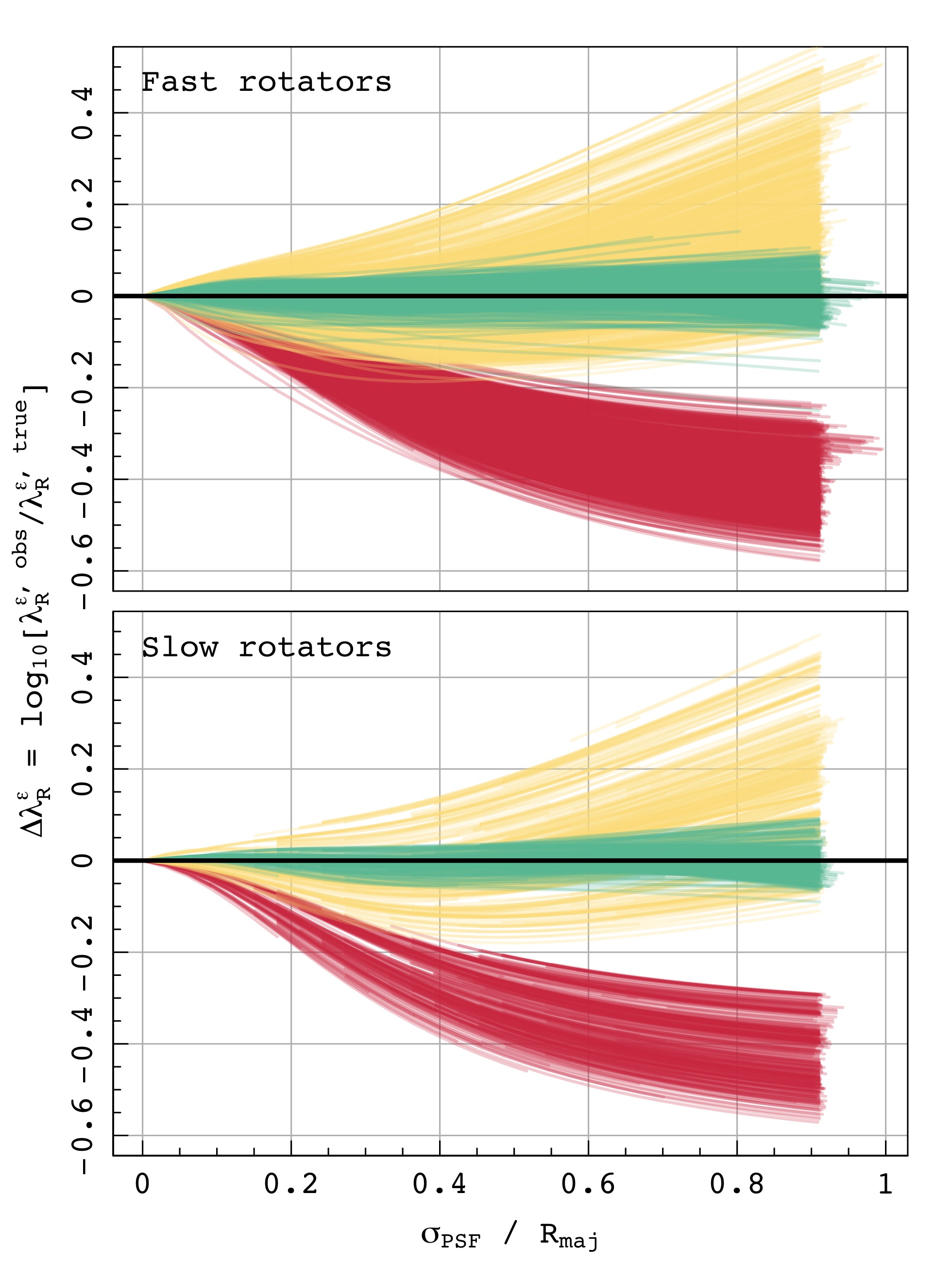}
    \caption{As in Figure \ref{fig:hf_correction} for FR measurements of $\lambda_R^{\varepsilon}$ (above), and for SR observations (below). We show this for the elliptically measured spin parameter \lR{}$^{\varepsilon}$. Red lines show the trends before correction, orange lines demonstrate following the application of the G18 correction within the bounds of its design, yellow lines show G18's correction for all other data and green lines show the effect of applying the corrections presented in this work.}  
    \label{fig:hf_reff_elr}
\end{figure}

Hence, following the methods described in Section  \ref{sec:method-correct}, we derived a second correction, this time using the $\lambda_R$ measured using the elliptical radii measure. This alternative correction is shown in Equation \ref{eq:elR_corr}. As before, we found that the HAR algorithm produces the lowest value of intrinsic scatter $\sigma(\Delta \lambda_{R}^{\varepsilon} = 0.04905$). The resulting equation is very similar to Equation \ref{eq:lr_corr}, but with a logarithmic dependence on ellipticity. In Figure \ref{fig:hf_reff_elr}, we show the effect of this correction on measurements of $\lambda_R^{\varepsilon}$.

\begin{center}
    \fbox{\begin{minipage}{.9\columnwidth}
\begin{equation} 
    \Delta \lambda_{\varepsilon-R}^{\text{corr}} = f\left(\frac{\sigma_{\text{PSF}}}{\text{R}_{\text{maj}}}\right) + \left(\frac{\sigma_{\text{PSF}}}{\text{R}_{\text{maj}}}\right) \times f(\varepsilon, n, \text{R}_{\text{eff}}^{\text{fac}}), \label{eq:elR_corr}
\end{equation}

\noindent where,
\begin{align}
    \begin{split}
    f\left(\frac{\sigma_{\text{PSF}}}{\text{R}_{\text{maj}}}\right)^{\Delta \lambda_{R}^{\varepsilon}} = \frac{7.49}{1 + e^{[4.01 \left(\frac{\sigma_{\text{PSF}}}{\text{R}_{\text{maj}}}\right)^{1.57} + 2.84]}} - 0.41,
    \end{split} \nonumber \\ 
    \begin{split}
    f(\varepsilon, n, \text{R}_{\text{eff}}^{\text{fac}})^{\Delta \lambda_{R}^{\varepsilon}} = [0.02 \times \text{log}_{10}(\varepsilon)] \\ - [0.19 \times \text{log}_{10}(n)] \\ - [0.13 \times  \text{R}_{\text{eff}}^{\text{fac}}] \\ + 0.28.\nonumber
    \end{split}
\end{align}
\end{minipage}
}
\end{center}

\begin{figure}
    \centering
	\includegraphics[width=0.75\columnwidth]{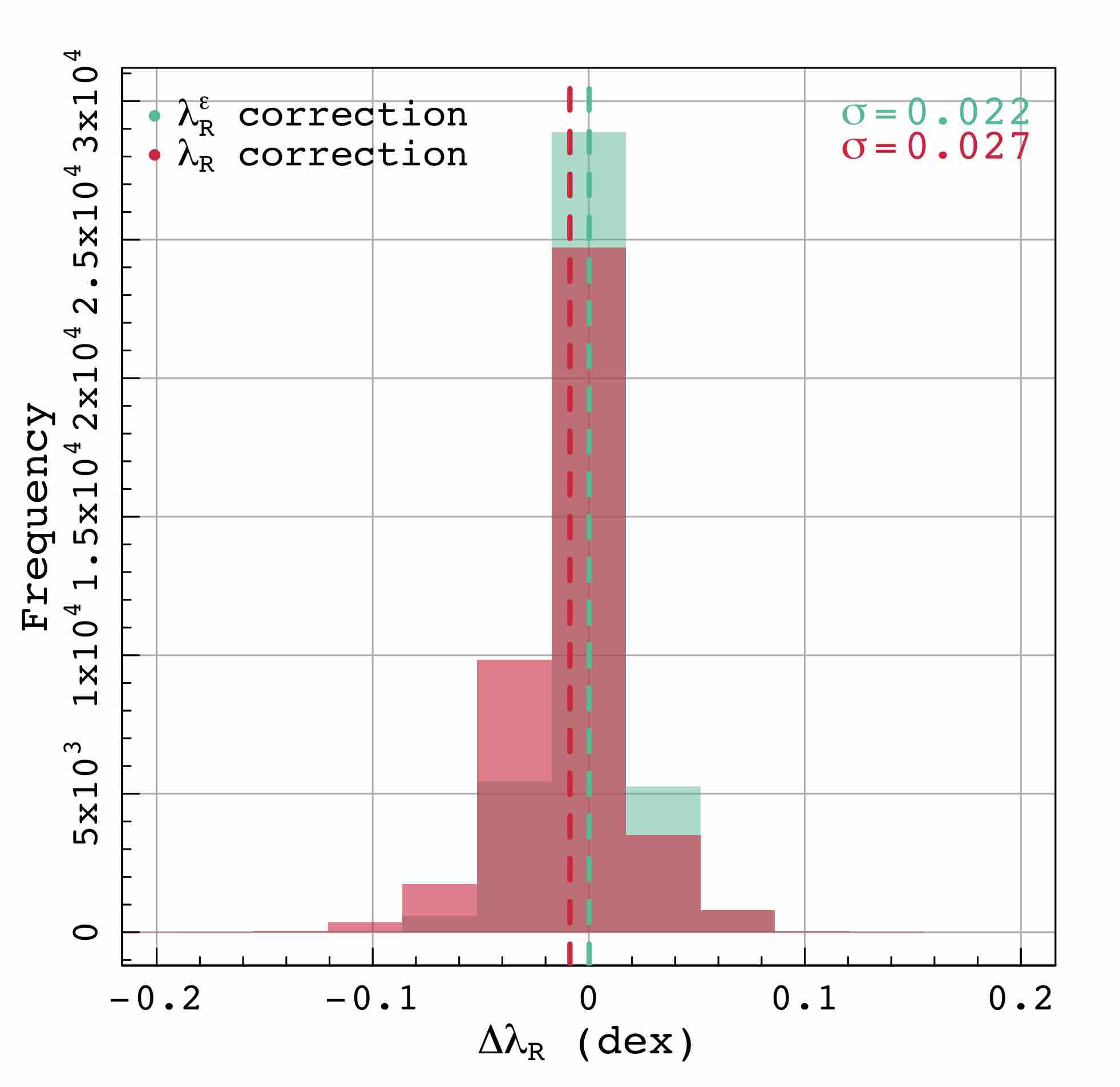}
    \caption{Showing the distributions of corrected values of $\Delta\lambda_R^{\varepsilon}$ when using Equation \ref{eq:elR_corr} (green) vs. Equation \ref{eq:lr_corr} (red). While the means of each histogram both lie close to zero, the red histogram is skewed much more than the green towards under-corrected values.}  
    \label{fig:residuals_elr}
\end{figure}

Given the two versions of this correction, we further tested how well the \lR{} version (Equation \ref{eq:lr_corr}) performed when using it to correct $\lambda_R^{\varepsilon}$, rather than Equation \ref{eq:elR_corr}. The results of this investigation are shown in Figure \ref{fig:residuals_elr}. While both corrections will bring measurements back towards the true value, when using Equation \ref{eq:lr_corr} on $\lambda_R^{\varepsilon}$, values will tend to be under-corrected. 

Plotting the $\lambda_R^{\varepsilon}$ measurements against \vsigma{}, as in Section \ref{sec:kappa}, we find that we recover smaller values of $\kappa \sim 0.91 - 0.95$ (in line with \cite{vandeSande2017TheSurveys} below unity). The effects of seeing on this definition of $\kappa$ show similar trends to Figure \ref{fig:kappa}, but are less severe, especially at \Reff{} where the distribution with seeing is almost flat. This seems to be because the increasing seeing moves you along the curve, rather than off the curve. 

\section{Comparing \texorpdfstring{$N$}{N}-body and hydro-dynamical treatments for flux}
\label{sec:app_flux}

Throughout this work, we have used $N$-body systems to derive our correction. These models have been observed using \simspin{} where the mass of each particle has been converted into a relative flux using a mass-to-light ratio. For the \eagle{} systems, we have followed the prescription laid out in \cite{Trayford2015ColoursSimulation}, and used SED modelling to calculate the flux of each particle according to the age and metallicity. 

To verify that this difference does not cause an impact on the measured results. Here, we took one of the high resolution FR galaxies, \texttt{GalaxyID = 14202037}, and observed it through \simspin{} using the mass-to-light prescription used on the $N$-body systems. We have taken 5758 \lR{} observations at the same range of seeing conditions, inclination and measurement radii as the original sample. We have then plotted these alongside the equivalent hydro-dynamical SED measurements. 

\begin{figure}
	\includegraphics[width=\columnwidth]{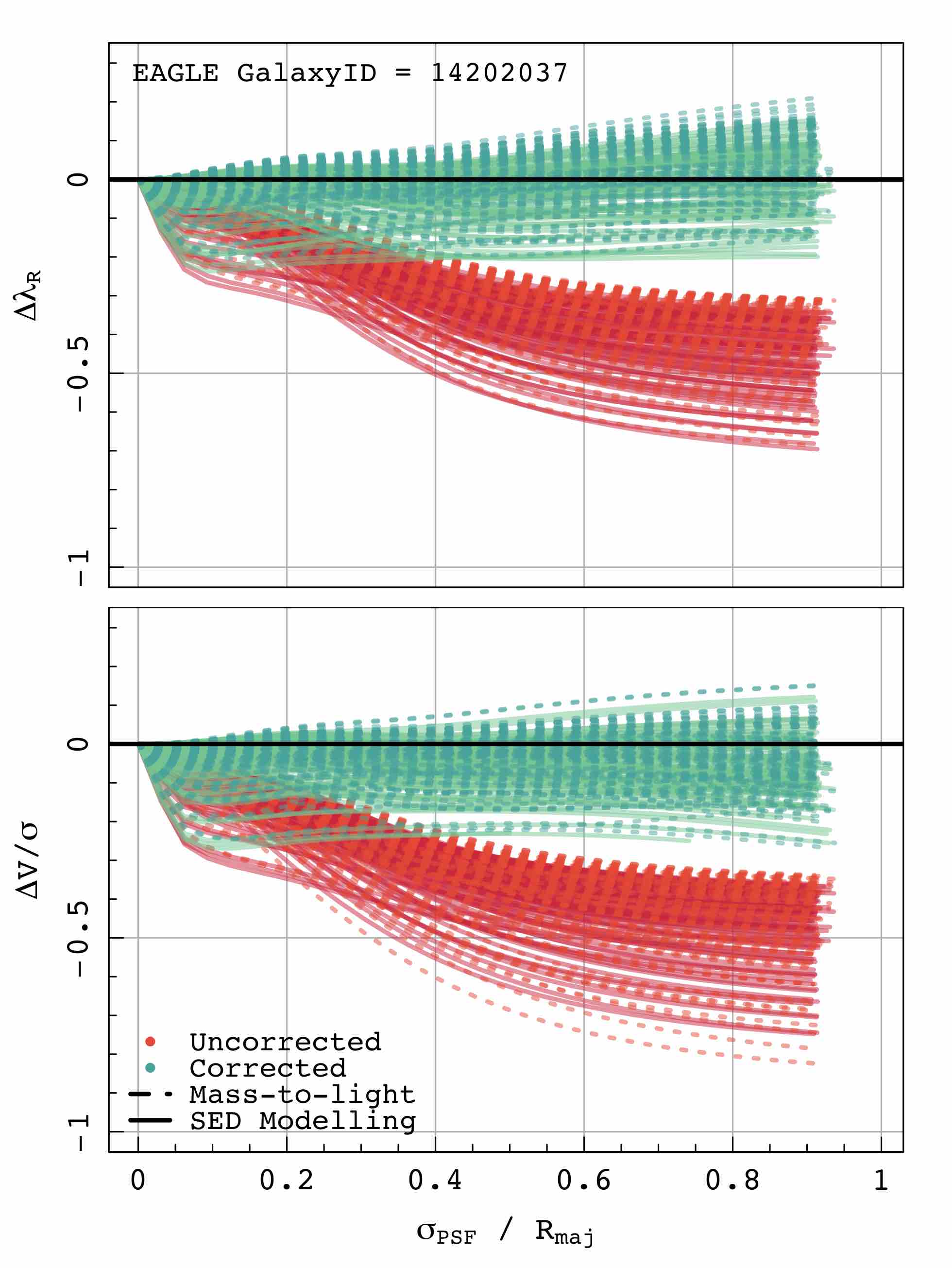}
    \caption{The tracks of $\Delta$\lR{} and $\Delta$\vsigma{} for observations of the \eagle{} galaxy \texttt{GalaxyID = 14202037} made using the two treatments of flux within \simspin. The mass-to-light treatment used for the N-body galaxies is shown in the dashed lines while the full SED modelling used for the hydro-dynamical galaxies is shown in the solid lines. Red tracks show the uncorrected measurements and green show tracks following the correction presented in this paper.}
    \label{fig:flux_comp}
\end{figure}

There is very little difference between the two treatments, as can be seen in Figure \ref{fig:flux_comp}. The following statistics are presented as the median of each distribution with the \nth{16} and \nth{84} percentiles below and above respectively ($\nu_{16\tiny{\text{th}}}^{84\tiny{\text{th}}}$). The residual values are \corrdlr{} $\sim -0.011_{-0.057}^{+0.080}$ and \corrdlr{} $\sim 0.012_{-0.089}^{+0.030}$ for the SED and mass-to-light treatments respectively. Overall, this difference is within the standard error such that the analysis made for the \eagle{} systems in Section \ref{sec:res_eagle} are valid.

\section{Examining the effect of spatial sampling}
\label{app:ch5_spaxels}
As discussed in Section \ref{sec:FR/SR_class}, the effects of spatial sampling and observational seeing are strongly related. In order to separate these effects, we have maintained a constant spatial sampling throughout the development of the correction and subsequent tests. However, it is important to assess how sensitive these corrections are to the spatial sampling of the observation.

In order to examine this, we have taken a selection of the $N$-body galaxies in Table \ref{tab:sims} (the S0, Sb and Sd galaxies at each concentration, giving a sample of 9 models) and re-measured the kinematic parameters, \lR{} and \vsigma{}, at a variety of different spatial samplings. We did this by observing the galaxy from different distances, evaluating each galaxy at 7 different sizes within the aperture and 10 inclinations, corresponding to a further 630 observations. We define the spatial pixel sampling (SPS) as a dimensionless quantity that represents the area of the pixels within the measurement radius. This has been parameterised as,
\begin{equation}
    \text{SPS} = (\pi R_{\text{maj}}^2 \sqrt{1 - \varepsilon^2}) / \delta_{\text{spaxel}}^2,
    \label{eq:SPS}
\end{equation}

where the SPS is dimensionless but related to the number of pixels within the measurement ellipse, given by the elliptical area of the measurement radius in arcsec divided by the square of the spaxel size of the IFU data cube ($\delta_{\text{spaxel}}$, i.e. SAMI's spatial pixels have a size 0.5''). In this definition, $\delta_{\text{spaxel}}$ is a constant defined by a survey's instrumentation and therefore cannot be changed to improve the SPS of the galaxy in question (i.e. you cannot re-bin your image with a greater number of pixels and hope to improve the corrected kinematics). In this test we have explored a range of samplings from SPS of 50 to 600.

\begin{figure}
	\includegraphics[width=\columnwidth]{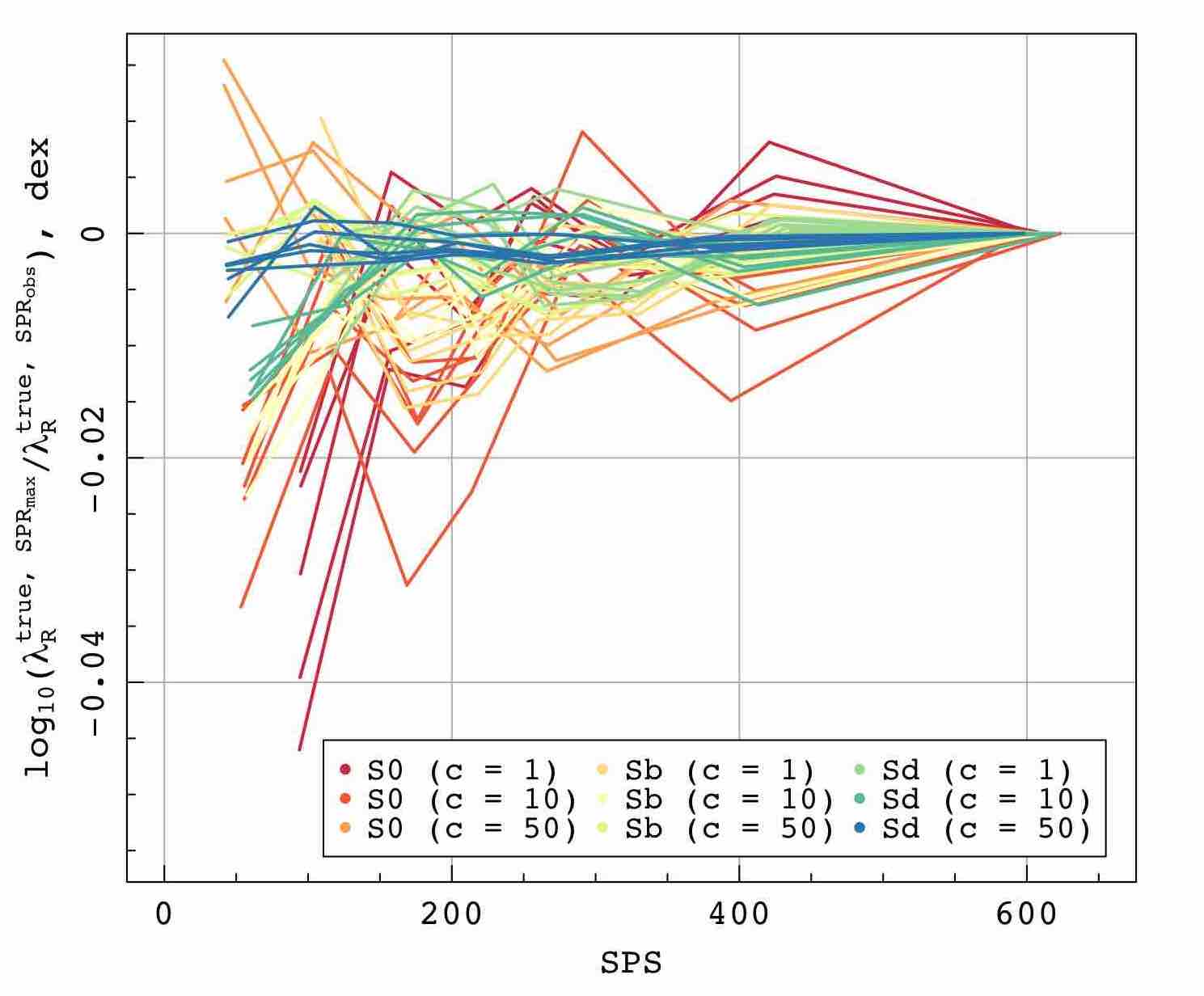}
    \caption{Demonstrating the effect of reducing the spatial sampling on kinematic measurements. No seeing conditions have been incorporated at this stage, such that the effects of seeing and spatial sampling can be disentangled. At SPS values below $\sim 250$, the measured kinematic values become quite noisy.}
    \label{fig:spaxel_limit}
\end{figure}

In Figure \ref{fig:spaxel_limit}, we show how the true kinematic value (the value that is measured at perfect seeing conditions) changes as the spatial sampling decreases. The difference is less that 0.01 dex when SPS is greater than $\sim 250$, but below this sampling the 1-$\sigma$ spread around zero becomes 0.05 dex. We note that the variation due to spatial sampling is considerable smaller than the impact of seeing when \psf{} $\sim$ 0.2. This spatial sampling threshold has a strong dependence on \sersic{} index, as demonstrated by the colour of the lines. More disk dominated galaxies are stable to lower samplings. Hence, the SPS will change our definition of $\lambda_R^{\text{true}}$. However, this does not mean that the correction presented in this work is not useful below this sampling, just that the quoted uncertainty in the corrected kinematics will become larger as the sampling becomes poorer.

\begin{figure}
	\includegraphics[width=\columnwidth]{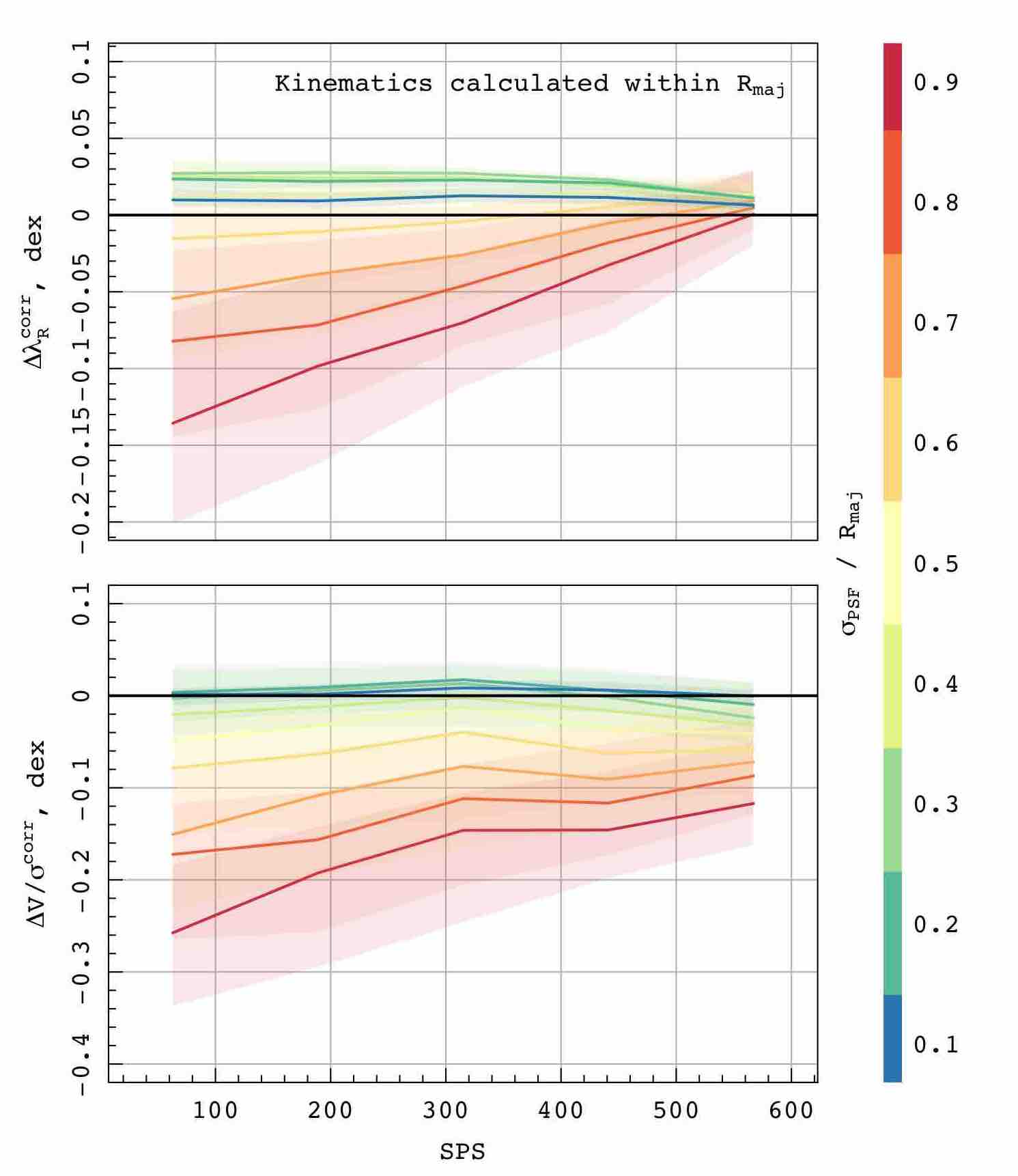}
    \caption{The relative difference between the corrected kinematics measured within 1 \Reff{} and their true value (\corrdlr{} = log$_{10}(\lambda_R^{\text{corr}}) -$ log$_{10}( \lambda_R^{\text{true}})$) across a range of different spatial pixels sampling. The colours represent the level of seeing, \psf. Each line shows the median, with the shaded region showing the 16th and 84th percentiles of the distribution.}
    \label{fig:spaxel_corr}
\end{figure}

In order to understand how the uncertainty of the corrected kinematics changes with spatial sampling, we have taken our 630 galaxy observations and applied seeing conditions at increments of 0.5'' from 0 to 15'' to give a total of 19530 observations. 
In Figure \ref{fig:spaxel_corr}, we demonstrate the effect of applying the seeing correction to each observation and plot the relative difference between the corrected kinematic value and the true value at that sampling. Colours demonstrate the level of seeing for each group of observations. Each line demonstrates the median with the shading showing the 16th and 84th percentiles about this value. By breaking up the distribution in this way, we can see that the application of this correction is valid down to the poorest SPSs considered, though the quoted uncertainty on these parameters will increase with the level of seeing. Hence, we caution against using this correction method for heavily seeing-dominated data with low spatial sampling. At the lowest spatial samplings, it would be sensible to apply a cut at \psf{} < 0.6 to avoid the significant rise in uncertainty that occurs in the kinematics of the most blurred observations. 

This is especially important to note for surveys like SAMI and MaNGA, whose sampling will often sit around SPS $\sim$ 250. However, following the standard data cuts made for SAMI DR2 in Section \ref{sec:res_DR2}, we find that while 459 galaxies sit with SPS < 250, only 2 of those systems sit in the maximum range 0.5 < \psf{} < 0.6. Hence, although DR2 contains a range of SPS, including values as low as 50, these observations are often at very good seeing conditions (\psf{} $\sim$ 0.2'') allowing the correction equation and its stated uncertainty to still be valid for these regimes.

\begin{figure}
	\includegraphics[width=\columnwidth]{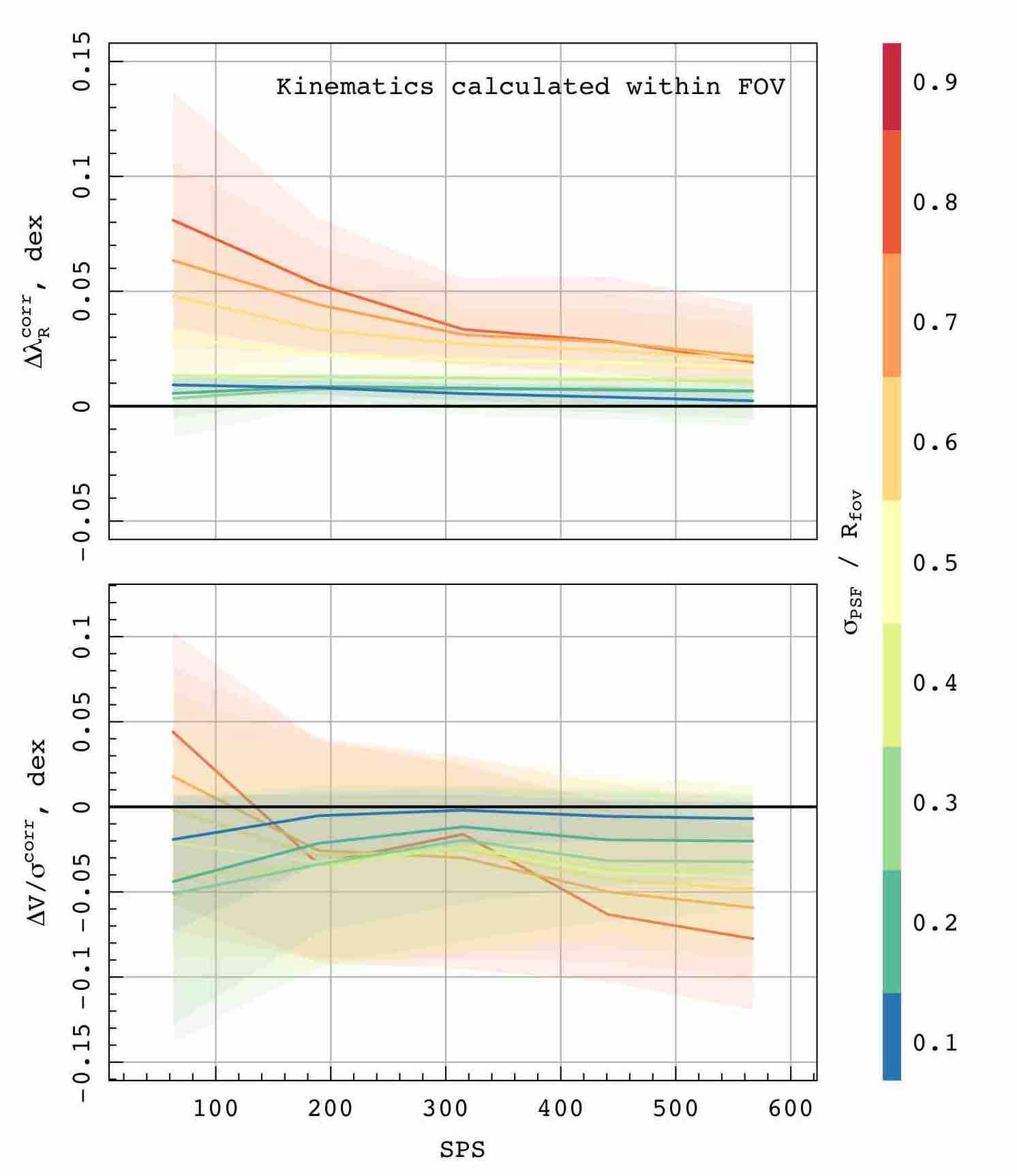}
    \caption{The relative difference between the corrected kinematic parameter as measured using all spaxels within the FOV and the true associated value across a range of spatial samplings. As in Figure \ref{fig:spaxel_corr}, each line shows the median with the 16th-84th percentiles indicated by the shaded region. Observations are coloured by the level of seeing, where we now consider the size of the PSF relative to the radius of the FOV (to replace \psf{} within the correction equations \ref{eq:lr_corr} and \ref{eq:vsig_corr})}.
    \label{fig:spaxel_corr_fov}
\end{figure}

Finally, we also examine the suitability of the idea proposed in \cite{DEugenio2013Fast0.183}, in which instead of using only a limited number of pixels within 1 \Reff, you use all pixels within the field of view to compute the kinematics. For each observation made at a reduced spatial sampling, we also compute the kinematics within an ellipse at the maximum extent of the field-of-view, noting the number of \Reff{} this occurs at. We can then use the \Reff{}-factor within the correction equations to remove the effects of seeing from this measurement. Considering these results as in Figure \ref{fig:spaxel_corr}, we show how the relative uncertainty in the corrected kinematics changes as a function of spatial sampling and seeing conditions in Figure \ref{fig:spaxel_corr_fov}. Again, the lines show the median value within each group of observations and the shaded regions show the 16th and 84th percentiles.

Because we are now correcting a measurement made at a greater radial extent, Figure \ref{fig:spaxel_corr_fov} does not mimic the same trend as the measurements made within 1 \Reff{} in Figure \ref{fig:spaxel_corr}. By measuring the kinematics using all available spaxels, there is likely to be a greater proportion of extraneous flux in the measured kinematic value and hence, it is more likely to be over-corrected. However, the corrected kinematics will have a more consistent uncertainty across the variety of spatial samplings considered. This technique also makes it easier to use measurements from the full variety of seeing conditions. Of course, the choice of using the full FOV to compute the kinematics will depend on the scientific enquiry being made. However, Figure \ref{fig:spaxel_corr_fov} shows that the correction presented in this work is also applicable in such cases due to the inclusion of the \Reff{}-factor parameter. By replacing the \psf{} value within Equations \ref{eq:lr_corr} and \ref{eq:vsig_corr} with $\sigma_{\text{PSF}} / R_{\text{FOV}}$ (in order to properly account for the level of seeing relative to the measurement radius), the correction works consistently well across a broader range of SPS. After measurements have been corrected using seeing at the maximum radial extent, the data can be corrected to a comparative value (i.e. 1 \Reff) using aperture corrections \citep[e.g.][]{vandeSande2017TheSurveys}. 

\section{Cut-out kinematic images of model galaxies}
\label{app:ch5_catalogue}

\begin{figure*}
\centering
	\includegraphics[width=0.85\textwidth]{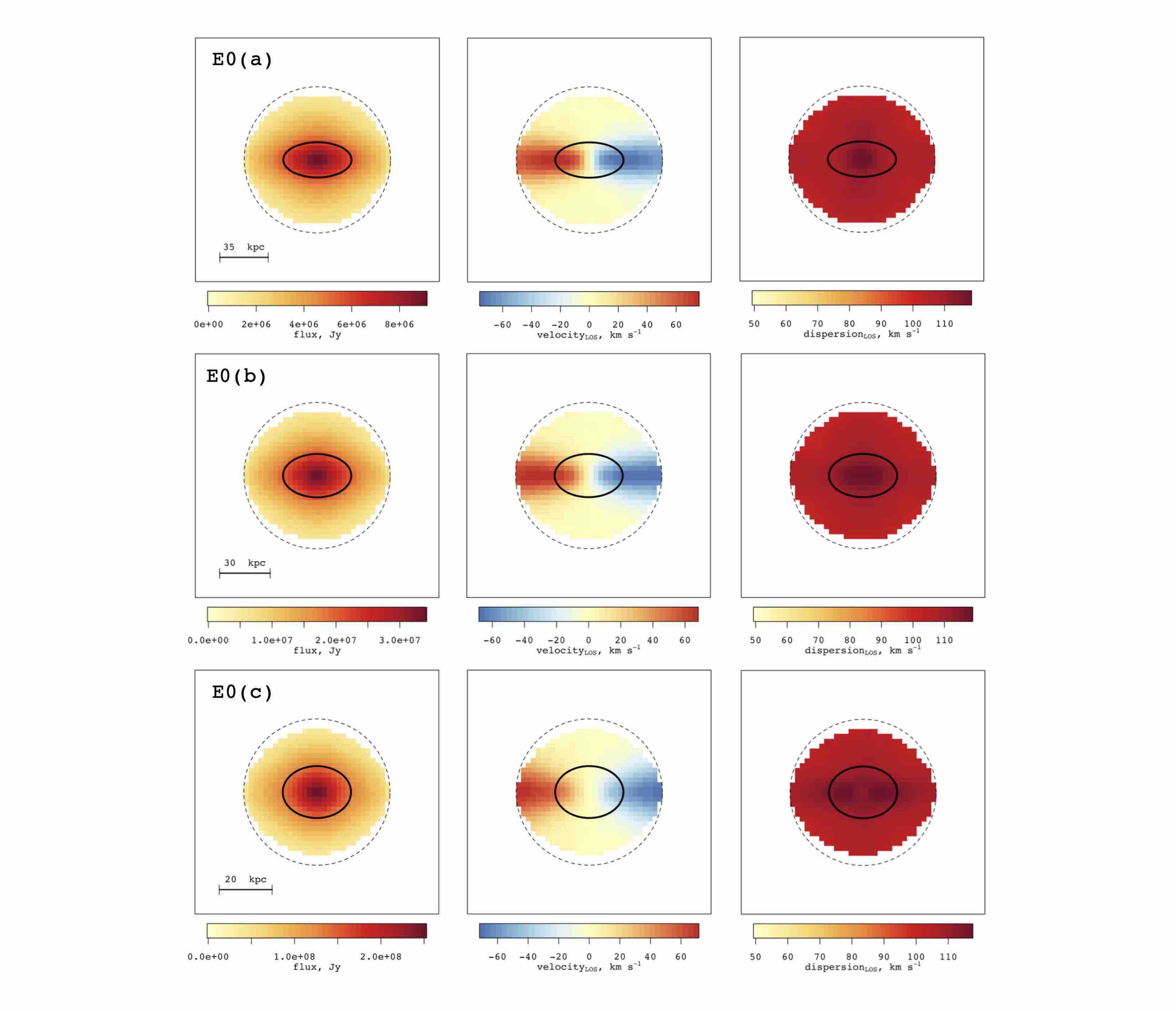}
    \caption{Cut-out images of the elliptical $N$-body galaxies observed using \simspin. Rows (a), (b) and (c) correspond to the first, second and third rows of E0 galaxies in Table 1. Each galaxy is projected at an angle of 70 degrees and has a PSF FWHM of 1''. Solid black lines show the 1 \Reff{} measurement radius of the system and the dashed black line demonstrates the full field of view of the SAMI observation.}
    \label{fig:ch5_nbody_E0s}
\end{figure*}

\begin{figure*}
\centering
	\includegraphics[width=0.85\textwidth]{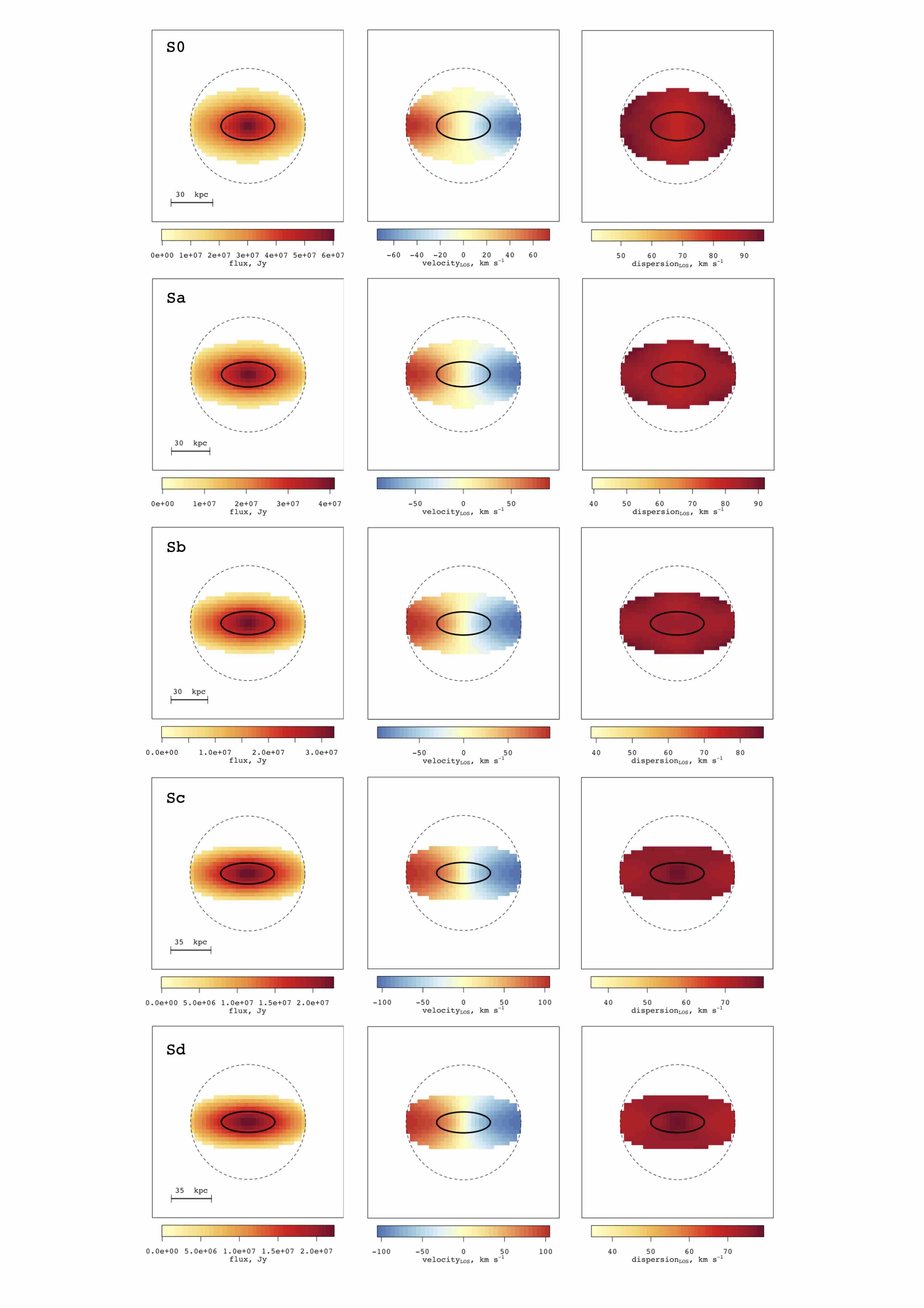}
    \caption{Cut-out images of the $c = 1$ $N$-body galaxies observed using \simspin. Each galaxy is projected at an angle of 70 degrees and has a PSF FWHM of 1''. Solid black lines show the 1 \Reff{} measurement radius of the system and the dashed black line demonstrates the full field of view of the SAMI observation.}
    \label{fig:ch5_nbody_c1}
\end{figure*}

\begin{figure*}
\centering
	\includegraphics[width=0.85\textwidth]{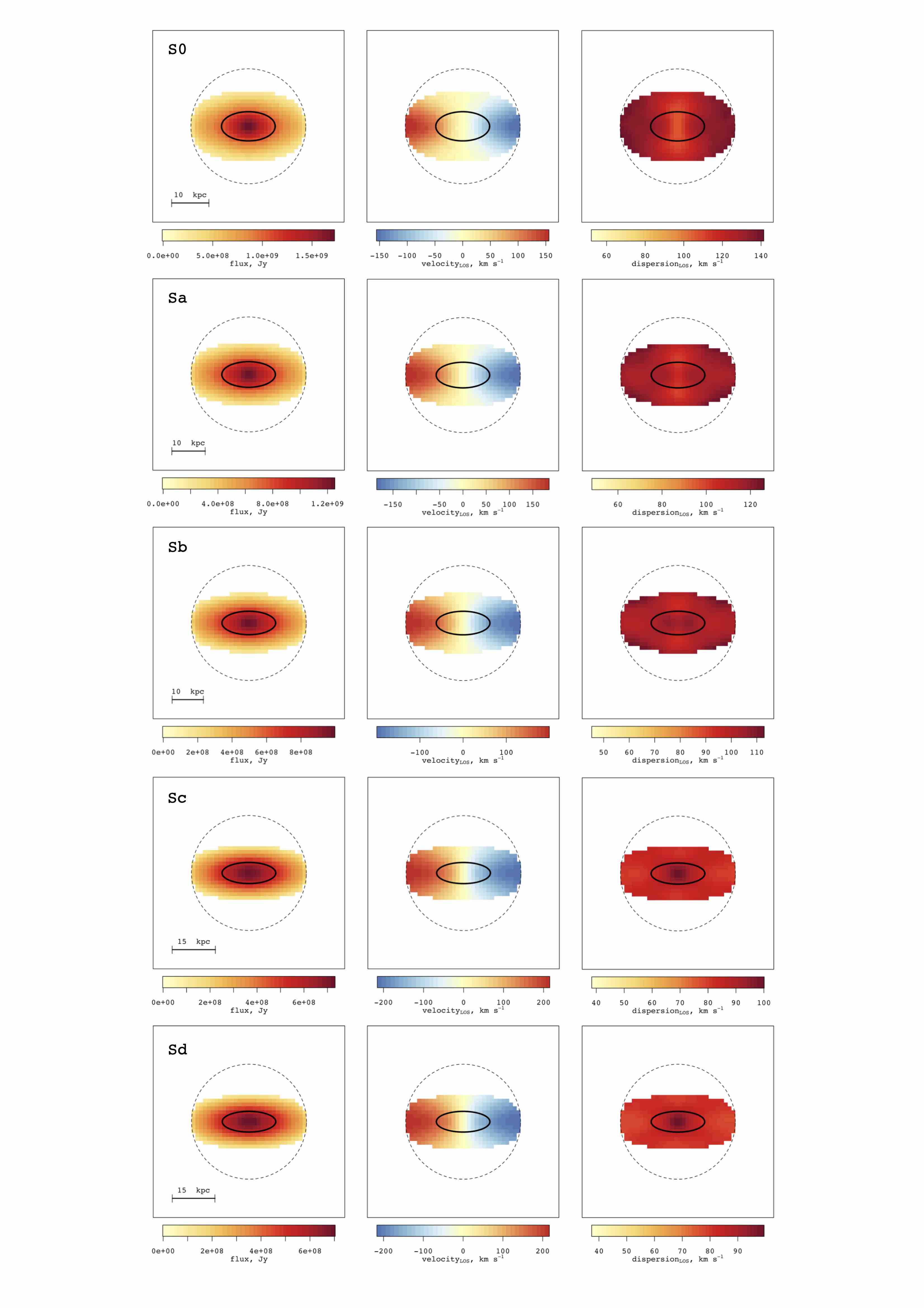}
    \caption{Cut-out images of the $c = 10$ $N$-body galaxies observed using \simspin. Each galaxy is projected at an angle of 70 degrees and has a PSF FWHM of 1''. Solid black lines show the 1 \Reff{} measurement radius of the system and the dashed black line demonstrates the full field of view of the SAMI observation.}
    \label{fig:ch5_nbody_c10}
\end{figure*}

\begin{figure*}
\centering
	\includegraphics[width=0.85\textwidth]{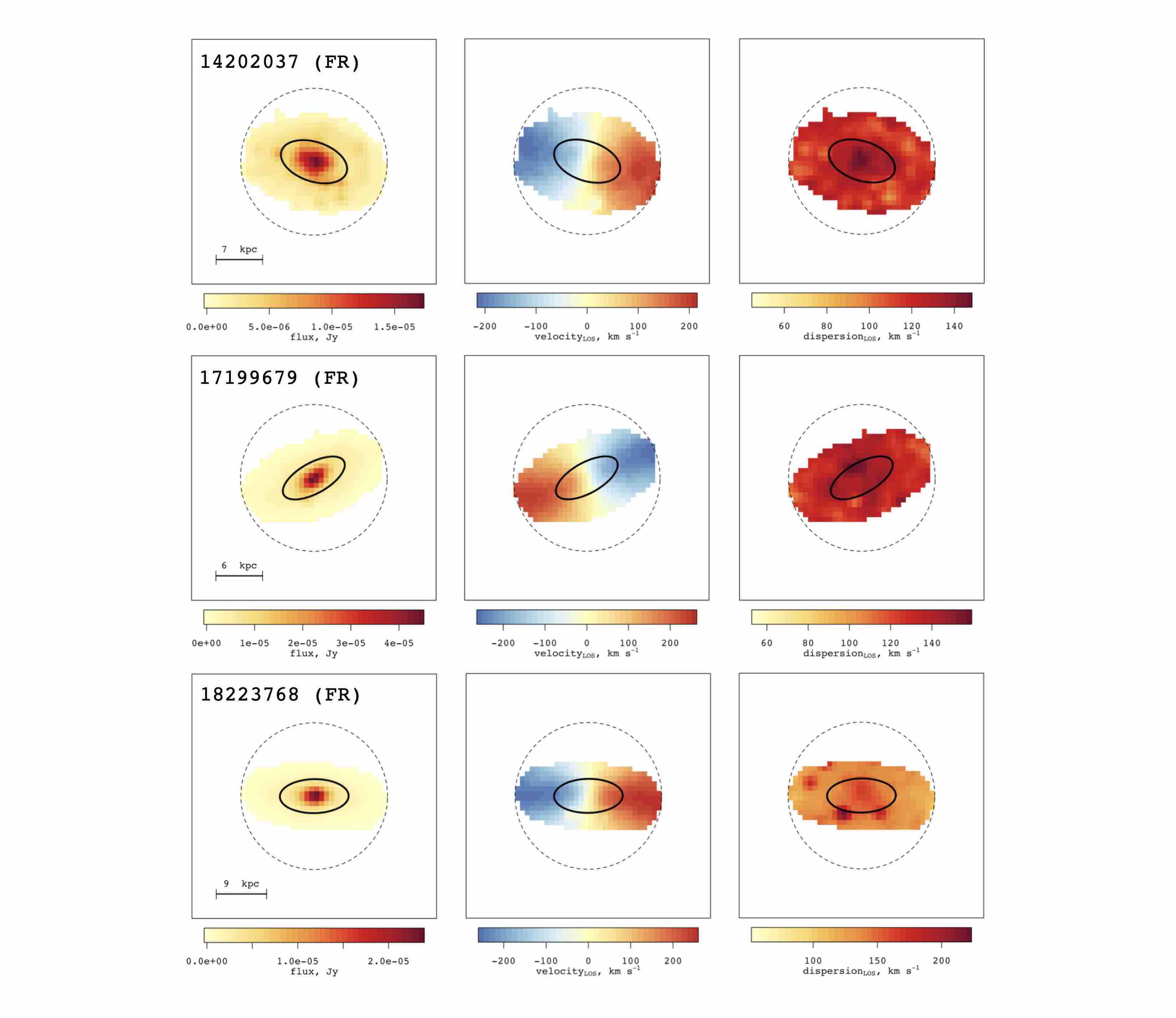}
    \caption{Cut-out images of the FR \eagle{} galaxies observed using \simspin. Each galaxy is projected at an angle of 70 degrees and has a PSF FWHM of 1''. Solid black lines show the 1 \Reff{} measurement radius of the system and the dashed black line demonstrates the full field of view of the SAMI observation.}
    \label{fig:ch5_eagle_FRs}
\end{figure*}

\begin{figure*}
\centering
	\includegraphics[width=0.85\textwidth]{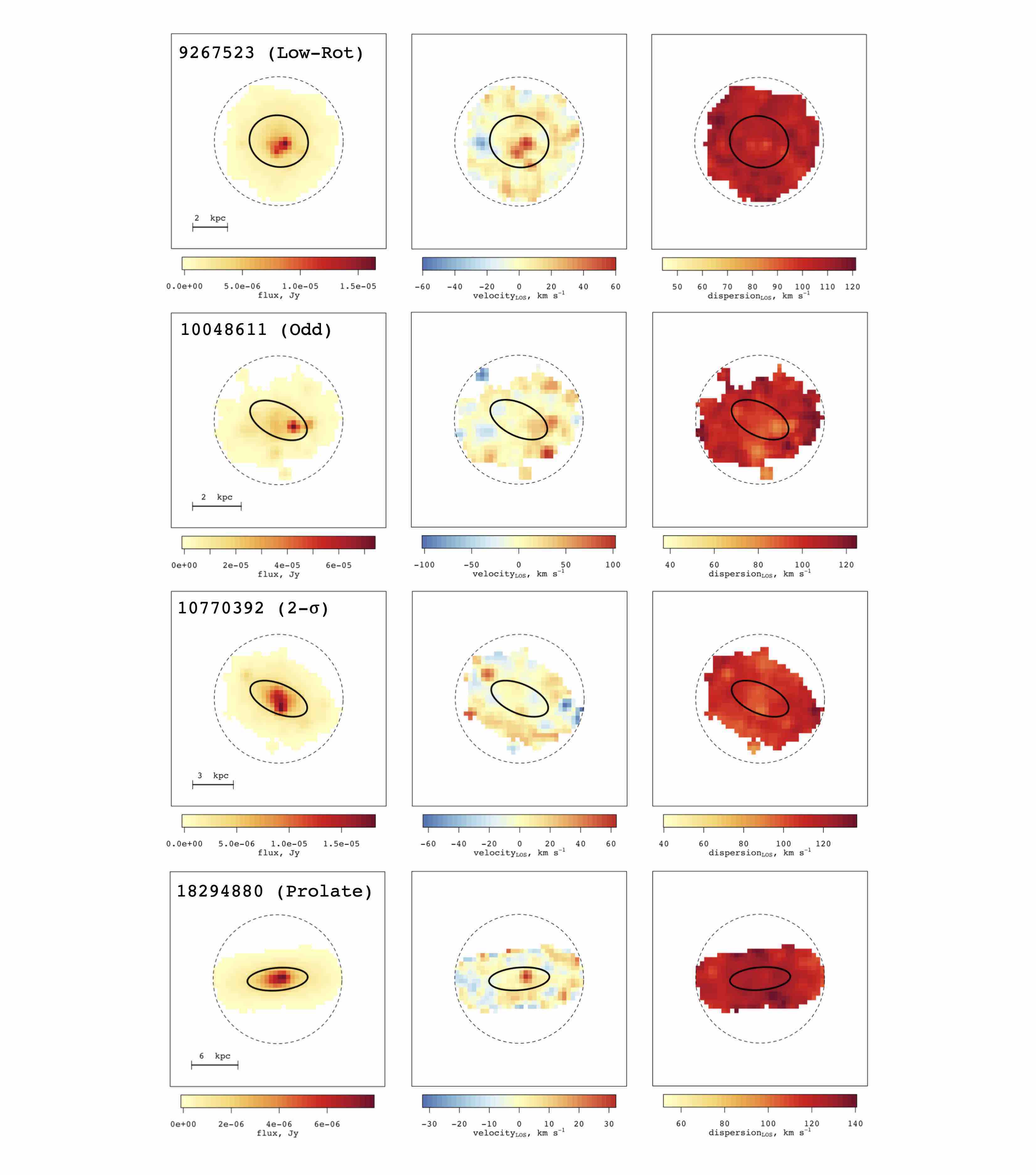}
    \caption{Cut-out images of the SR \eagle{} galaxies observed using \simspin. Each galaxy is projected at an angle of 70 degrees and has a PSF FWHM of 1''. Solid black lines show the 1 \Reff{} measurement radius of the system and the dashed black line demonstrates the full field of view of the SAMI observation.}
    \label{fig:ch5_eagle_SRs}
\end{figure*}

Here we present the kinematic maps generated for all of the galaxies examined in this paper. The elliptical systems are then shown in Figure \ref{fig:ch5_nbody_E0s}. In Figure \ref{fig:ch5_nbody_c1}, we show \textsc{Sami}-like images for the $c = 1$ model from each morphological type in the $N$-body catalogue, Figure \ref{fig:ch5_nbody_c10} shows the $c = 10$ models and Figure \ref{fig:ch5_nbody_c10} shows the $c = 50$ models as shown in Table 1. In Figure \ref{fig:ch5_eagle_FRs}, we demonstrate the \textsc{Sami}-like images created using \simspin{} for three FR galaxies in the \eagle{} sample, and in Figure \ref{fig:ch5_eagle_SRs} we show the remaining four SRs as shown in Table 2. The ``odd'' system is shown in the second row.

These images have all been produced with a PSF = 1'' and each galaxy is projected at an inclination of 70 degrees. For interested readers, we note that a dispersion minimum is found at the centre of these bulge-dominated, $N$-body galaxies, which is clearer in the models with higher concentrations. This is due to the distribution function of the Hernquist bulges \citep{Hernquist1990AnBulges} with which the models are built and is an entirely physical element of the projected LOS-velocity \citep{Baes2002TheModels}.

\begin{figure*}
\centering
	\includegraphics[width=0.85\textwidth]{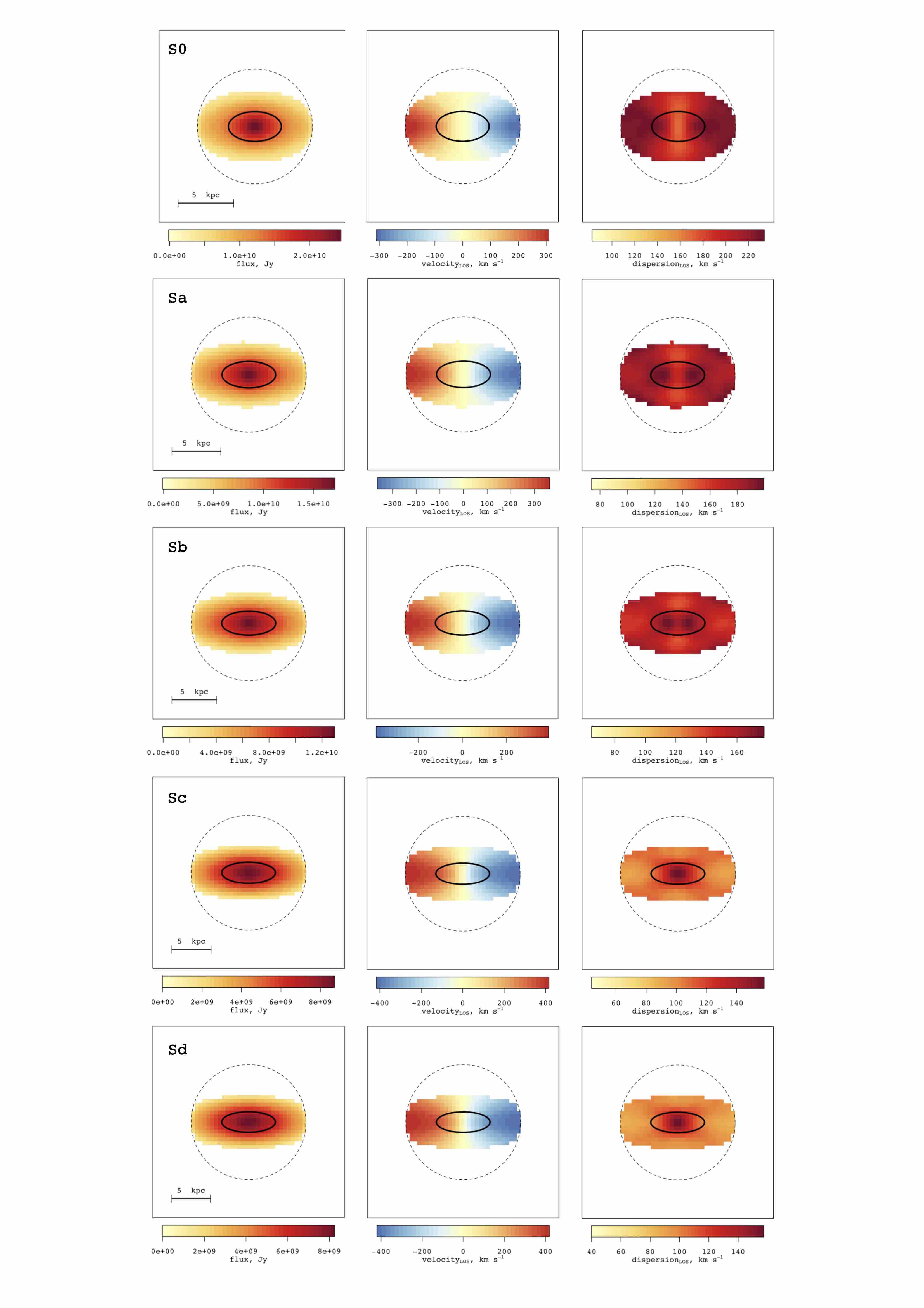}
    \caption{Cut-out images of the $c = 50$ $N$-body galaxies observed using \simspin. Each galaxy is projected at an angle of 70 degrees and has a PSF FWHM of 1''. Solid black lines show the 1 \Reff{} measurement radius of the system and the dashed black line demonstrates the full field of view of the SAMI observation.}
    \label{fig:ch5_nbody_c50}
\end{figure*}

%%%%%%%%%%%%%%%%%%%%%%%%%%%%%%%%%%%%%%%%%%%%%%%%%%

% Don't change these lines
%\bsp	% typesetting comment
\label{lastpage}
\end{document}